\documentclass[pre,aps,twocolumn,showpacs]{revtex4-1} %For final 
\newif\ifusebibtex
\usebibtexfalse

\newif\ifshowoutline
\showoutlinefalse

\newif\ifshowfigurelist
\showfigurelistfalse

\newif\ifshowpaper  %Needs to be true for the next few to work when true
\showpapertrue

\newif\ifshowmain
\showmaintrue

\newif\ifshowappendix
\showappendixtrue

\usepackage{epsfig}
\usepackage{psfrag}
\usepackage{amsmath}
\usepackage{amsfonts}
\usepackage{amssymb}
\usepackage[normalem]{ulem} %For \sout{}

\usepackage{graphicx}

\usepackage{color}
\definecolor{RED}{rgb}{1,0,0}
\definecolor{BLUE}{rgb}{0,0,1}
\definecolor{GREEN}{rgb}{0,1,0}
\definecolor{MAGENTA}{rgb}{1,0,1}
\newcommand{\red}[1]{{\color{RED} #1}}

\def\vc{\boldsymbol}
\def\mtx{\mathbf}
\def\mc{\mathcal}
\def\reals{\mathbb{R}}
\newcommand{\abs}[1]{\left| #1 \right|}
\newcommand{\norm}[1]{\left| #1 \right|}
\newcommand{\order}[1]{\mathcal{O}\left( #1 \right)}

\begin{document}
\title{Static Structural Signatures of Nearly Jammed Disordered and Ordered Hard-Sphere Packings: Direct Correlation Function}
\author{Steven Atkinson}
\affiliation{Department of Mechanical and Aerospace Engineering, Princeton University, Princeton, New Jersey 08544, USA}
\author{Frank H. Stillinger}
\affiliation{Department of Chemistry, Princeton University, Princeton, New Jersey 08544, USA}
\author{Salvatore Torquato}
\email{torquato@princeton.edu}
\affiliation{Department of Chemistry, Department of Physics, Princeton Center for Theoretical Science, Program of Applied and Computational Mathematics, Princeton Institute for the Science and Technology of Materials, Princeton University, Princeton, New Jersey 08544, USA}
\date{\today}

%=======================================================
% 
% Abstract
%

\begin{abstract}
%Motivate. 
%Static signatures.  
%Hyperuniformity. 
%Protocols.
%FCC. 
%Discontinuities...
%
The nonequilibrium process by which hard-particle systems may be compressed into disordered, jammed states has received much attention because of its wide utility in describing a broad class of amorphous materials.
While dynamical signatures are known to precede jamming, the task of identifying static structural signatures indicating the onset of jamming have proven more elusive.
The observation that compressing hard-particle packings towards jamming is accompanied by an anomalous suppression of density fluctuations (termed ``hyperuniformity'') has paved the way for the analysis of jamming as an ``inverted critical point'' in which the direct correlation function $c(\vc r)$, rather than the total correlation function $h(\vc r)$ diverges.
We expand on the notion that $c(\vc r)$ provides both universal and protocol-specific information as packings approach jamming.
By considering the degree and position of singularities (discontinuities in the $n$-th derivative) as well as how they are changed by the convolutions found in the Ornstein-Zernike equation, we establish quantitative statements about the structure of $c(\vc r)$ with regards to singularities it inherits from $h(\vc r)$.
These relations provide a concrete means of identifying features that must be expressed in $c(\vc r)$ if one hopes to reproduce various details in the pair correlation function accurately and provide stringent tests on the associated numerics.
We also analyze the evolution of systems of three-dimensional monodisperse hard-spheres of diameter $D$ as they approach ordered and disordered jammed configurations.
For the latter, we use the Lubachevsky-Stillinger (LS) molecular dynamics and Torquato-Jiao (TJ) sequential linear programming algorithms, which both generate disordered packings, but can show perceptible structural differences.
We identify a short-ranged scaling $c(\vc r) \propto -1/r$ as $r \rightarrow 0$ that accompanies the formation of the delta function at $c(D)$ that indicates the formation of contacts in all cases, and \ show that this scaling behavior is, in this case, a consequence of the growing long-rangedness in $c(\vc r)$, e.g., $c \propto -1/r^2$ as $r \rightarrow \infty$ for disordered packings.
At densities in the vicinity of the freezing density, we find striking qualitative differences in the structure factor $S(\vc k)$ as well as $c(\vc r)$ between TJ- and LS-generated configurations, including the early formation of a delta function at $c(D)$ in the TJ algorithm's packings, indicating the early formation of clusters of particles in near-contact.
Both algorithms yield structure factors that tend towards zero in the low-wavenumber limit as jamming is approached.
Correspondingly, we observe the expected power-law decay in $c(\vc r)$ for large $r$, in agreement with previous theoretical work.
Our work advances the notion that static signatures are exhibited by hard-particle packings as they approach jamming and underscores the utility of the direct correlation function as a sensitive means of monitoring for the appearance of an incipient rigid network.
\end{abstract}
\maketitle

%=======================================================
% 
% Outline
%

%showoutline

%=====================================================================
%
% Here's the actual paper
%
%=====================================================================
\ifshowpaper

%%%%%%%%%%%%%%%%%%%%%%%%%%%%%%%%%%%%%%%%%%%%%%%%%%%
%%                INTRODUCTION                   %%
%%%%%%%%%%%%%%%%%%%%%%%%%%%%%%%%%%%%%%%%%%%%%%%%%%%
%
% Things that are missing:
%
% ?
%
%
% Define packing
% Define jamming
% As jamming as approached, we have emergent long-range order (hyperuniformity)
%   The direct correlation function shows this as:
%     Divergence at the origin like r^-1 (in 3D)
%     MRJ: s(k) linear -->c(r) should decay as r^-2 (long-r)
%     FCC: S(k)~const (NOT quadratic); quadratic in k after some k_min -> c(r) decays as r^(-1) INSIDE of some r_max which is on the order of unity.
% Can this be linked to the incipient backbone?

\section{Introduction}
\label{sec:intro}

% Define packing
Packings of hard particles in $d$-dimensional Euclidean space $\reals^d$ have been used ubiquitously as a powerful model to describe many-body systems such as liquids, glasses, colloids, granular materials, particulate composites, and biological systems, among others \cite{Bernal_1965,Zallen_1983,Cates_1999,Torquato_2001,Torquato_2003_Breakdown,Torquato_2003_Hyperuniform,Zachary_2009,Liu_2003,Manoharan_2003,Makse_2004,Xu_2005,Zohdi_2006,Majmudar_2007,Gevertz_2008,Laso_2009,Parisi_2010,Schweizer_2010,Liu_2011,Charbonneau_2012,Dagois-Bohy_2012,Gillman_2013,Klatt_2014,Bowles_2015,Chaikin_2015,Schroder-Turk_2015,Chakraborty_2016}.
In three dimensions, the venerable hard-sphere model is particularly useful, owing to its mathematical simplicity and the rich diversity of equilibrium and non-equilibrium behavior that it exhibits.

%Jamming and inverted c.p.
It has been shown that bringing hard-particle packings towards jamming (roughly speaking, mechanical stability) is accompanied by an anomalous suppression of long-range density fluctuations \cite{Silbert_2009,Zachary_2011_PRE1,Zachary_2011_PRE2,Hopkins_2012,Durian_2016,Atkinson_2016_Slowdown}---a phenomenon known as ``hyperuniformity'' \cite{Torquato_2003_Hyperuniform,Zachary_2009}.
A many-particle system is hyperuniform if the structure factor $S(\vc k)$ (trivially related to the Fourier transform of pair statistics in direct space) tends to zero in the limit that the wavenumber $\norm{\vc k}$ tends to zero.
Hyperuniformity may be conceptualized as an ``inverted critical point'' in which the direct correlation function $c(\vc r)$, which is defined through the Ornstein-Zernike integral equation for a system with number density $\rho$:
\begin{equation}
h(\vc r_{12}) = c(\vc r_{12}) + \rho \int_{\reals^d} h(\vc r_{23}) c(\vc r_{13}) d \vc r_3,
\label{eqn:OZ}
\end{equation}
 becomes long-ranged, i.e., its volume integral diverges \cite{Torquato_2003_Hyperuniform}.
This is to be contrasted with the usual thermal critical point (e.g., liquid-vapor or Curie critical points) in which the total correlation function $h(\vc r)$ (rather than $c(\vc r)$) becomes long-ranged. 
Accordingly, a static length scale, obtained from the Fourier transform of $c(\vc r)$, $\xi = [-\tilde{c}(\vc k=\vc 0)]^{1/d}$, grows as a system approaches a hyperuniform state and ultimately diverges at this critical state. 
In hard-particle packings, this occurs as jamming is approached \cite{Hopkins_2012}, and a similar analysis can be used to obtain meaningful information about the nature of glassy states of particles with soft interaction potentials \cite{Marcotte_2013}. 
Because of this, it has been an intriguing prospect to investigate disordered hyperuniform systems by adapting standard tools used to investigate critical phenomena.

%OZ and PY; previous work.
At the same time, the direct correlation function has proven to be a fruitful starting point for efforts to characterize the structure of disordered systems such as simple liquids since the pair statistics may be obtained through Eq.\ (\ref{eqn:OZ}) \cite{Hansen_1990}. 
Physically, the equation suggests that pair statistics between any two particles may be decomposed into a ``direct'' contribution encoded in $c(\vc r)$ as well as an ``indirect'' contribution mediated through ``chains'' of particles, expressed mathematically through the convolution between $h(\vc r)$ and $c(\vc r)$. 
For systems with suitably well-behaved interactions, one may equivalently think of $c(\vc r)$ as describing the linear response of a system to a perturbation in an externally applied potential field \cite{Hansen_1990}.

%Hopkins recent history
It has been observed that maximally random jammed (MRJ) hard-sphere packings, which constitute the most disordered configurations as measured by some scalar order metric subject to the constraint of jamming and isostaticity \cite{Torquato_2000,Torquato_2010}, are hyperuniform and exhibit a nonanalytic linear behavior in the structure factor for low $k$, namely, $S \propto k$ \cite{Donev_2005_Unexpected}. 
Hopkins et al.\ \cite{Hopkins_2012} studied the behavior of very large ($N=10^6$) sphere packings produced by the Lubachevsky-Stillinger event-driven molecular dynamics algorithm \cite{Skoge_2006} under rapid compression so as to study the approach to the MRJ-like states at densities well above the freezing density and close to jamming. 
They found evidence that the nonanalytic linear behavior in $S(\vc k)$ was evident considerably in advance of jamming, and that upon further compression, the extrapolated value at the origin tended towards zero, implying that corresponding long-ranged behavior in $c(\vc r)$ might be observable for this protocol.
Their computations for the Fourier-transformed direct correlation function showed $\tilde{c}$ extending towards negative infinity near the origin as the packings were compressed, supporting this prediction.

%What we want to find
In the current work, we further develop the view that one may find static, structural precursors to jamming in hard-particle systems. 
Because $c(\vc r)$ is known to generally possess a qualitatively simpler functional form while still encoding the complete pair statistics of the system, we will focus primarily on the signatures therein, paying particular attention to features that point towards the development of an incipient contact network and hyperuniform density fluctuations (i.e.\ long-rangedness in $c(\vc r)$). 
Moreover, $g_2(\vc r)$ is known to possess various singularities at jamming (e.g., a Dirac delta function at contact, discontinuity at a distance of two diameters), and we determine to what extent these features are inherited by $c(\vc r)$.

However, one should also bear in mind that different packing protocols will tend to produce different ensembles of disordered jammed states \cite{Jiao_2011}.
In this paper, we will also bring attention to qualitative differences in protocols' {\it approach} to their jammed states.
Additionally, it is nontrivial to ensure that standard protocols approach properly-jammed states and avoid becoming stuck in unstable mechanical equilibria. 
This has been found to be related to a ``critical slowing down'' that becomes of practical concern for large systems \cite{Atkinson_2016_Slowdown}. 
Therefore, we carry out our current investigation as if our systems are indeed jammed and hyperuniform with the important caveat that 
this is more difficult to do with high precision in practice
than previously thought. 
If one had a protocol that were able to produce better-jammed packings, we expect that the packings would possess stronger structural signatures consistent with hyperuniformity.

%What we do
To this end, we analyze computer-generated packings of monodisperse hard-spheres of diameter $D$ created by the LS algorithm as well as the Torquato-Jiao (TJ) sequential linear programming algorithm \cite{TJ_2010}. 
We consider these two algorithms since (i) they are both known to generate highly disordered packings under suitable conditions, but also because (ii) the jammed states they produce possess considerable differences in their macroscopic properties, including density, rattler fraction, and degree of order as measured by various standard order metrics \cite{Atkinson_2013}. 
By investigating multiple protocols that differ considerably, we seek to discern what features are protocol-dependent, and which are in common to a diversity of MRJ-like states.
%We find that the two protocols display significantly different qualitative behavior as their packings are compressed through intermediate densities (i.e.\ in the vicinity of the freezing density and above, but still far from jamming), but that they share emergent hyperuniform signatures as they approach jamming.
We also briefly consider the behavior of the hard-sphere FCC crystal, conjectured to be the equilibrium phase \cite{Mau_1999,Frenkel_2003} at packing fractions $\phi \in [0.55,\phi_{FCC})$ along the solid branch ending at close packing with a packing fraction of $\phi_{FCC} = \pi/\sqrt{18}$. 
This case provides valuable information about jamming under an arguably more well-behaved setting, where one need not worry about metastability, and hyperuniformity may be approached to arbitrary numerical precision with minimal practical issues.

%What we find
We observe that the TJ and LS protocols exhibit markedly different qualitative behavior in $c(\vc r)$, even at packing fractions far from jamming. 
Specifically, we find that the direct correlation of packings produced by the TJ algorithm exhibit signs of a delta function at $r=D$ at packing fractions below the freezing density $\phi \approx 0.494$, whereas features in $c(\vc r)$ exhibited by LS for $r>D$ are substantially more subtle up until much higher densities. 
With the development of the expected delta function at $c(r=D)$, we observe a concomitant development of a dominant $-1/r$ scaling for $c(r<D)$ as $r \rightarrow 0$.
We show using theoretical arguments that one can be predict this numerical observation.
Interestingly, we observe the power-law scaling $c(\vc r) \propto -1/r^2$ in the limit $r \rightarrow \infty$ that is a consequence of the linear trend in $S(\vc k)$ for small $k$. 
Observing this behavior is difficult in practice because it requires an accurate measurement of $S(\vc k)$ for small wavenumbers, which requires that one consider large packings. 
Our work advances the notion that static signatures are exhibited by hard-particle packings as they approach jamming and underscores the utility of the direct correlation function as a sensitive means of monitoring for the appearance of an incipient rigid network.

%The rest
The remainder of the paper is organized as follows: in Sec.\ \ref{sec:OZJamming}, we discuss some relevant analytical results pertaining to the structure of disordered hard-sphere packings through $c(\vc r)$ and the Ornstein-Zernike equation.
In Sec.\ \ref{sec:Inheritance}, we discuss quantitatively the manner by which $c(\vc r)$ inherits singularities from $g_2(\vc r)$.
In Sec.\ \ref{sec:Scaling}, we review some known facts regarding the critical scaling behavior expected for systems approaching hyperuniformity. 
In Sec.\ \ref{sec:Methods}, we review the protocols that we use to generate nearly-jammed hard-sphere packings. 
In Sec.\ \ref{sec:Results}, we present the structure factor and direct correlation functions of our ordered and disordered packings as they approach jamming and point out emergent static structural features exhibited by each. 
Conclusions and discussion are presented in Sec.\ \ref{sec:Conclusion}.

%%%%%%%%%%%%%%%%%%%%%%%%%%%%%%%%%%%
%%            THEORY             %%
%%%%%%%%%%%%%%%%%%%%%%%%%%%%%%%%%%%

\section{Ornstein-Zernike Equation and Jamming}
\label{sec:OZJamming}

We begin by reviewing a number of relationships between the standard pair statistical descriptors of point processes with the goal of relating the direct correlation function to other familiar statistical descriptors. 
We then proceed by reviewing theoretical progress that has been made in obtaining an accurate description of disordered hard-sphere systems by use of the direct correlation function, including shortfalls that persist with the current state of the art which point to the necessity for our present numerical investigations.

The structure factor is defined for a translationally-invariant system in $\reals^d$ as
\begin{equation}
S(\vc k) = 1 + \rho \tilde h(\vc k),
\label{eqn:Sk}
\end{equation}
where $\tilde h(\vc k)$ is the Fourier transform of the total correlation function. 
This is related to the scattering intensity $\mc S(\vc k)$, defined for a single system of $N$ particles within a fundamental cell under periodic boundary conditions as
\begin{equation}
\mc S(\vc k) = \frac{1}{N} \left| \sum_{j=1}^N e^{-i \vc k \cdot \vc r_j} \right|^2,
\label{eqn:SkCC}
\end{equation}
which includes forward scattering, i.e., $\mc S(\vc 0) \equiv N$.
This is to be contrasted with the definition of Eq.\ (\ref{eqn:Sk}), in which $S(\vc 0)$ is related to the volume integral of $h(\vc r)$. 
Apart from at $\vc k=\vc 0$, the scattering intensity is identical to the structure factor for a single configuration.
For an ensemble of periodic point configurations such (e.g. derived from the particles centers of our packings), the ensemble average of $\mc S(\vc k)$ is directly related to the structure factor $S(\vc k)$ via
\begin{equation}
\lim_{N,v_F \rightarrow \infty} \langle \mc S(\vc k) \rangle = (2 \pi)^d \delta(\vc k) + S(\vc k),
\end{equation}
where $\delta(\vc k)$ is the Dirac delta function and the limit being taken on $N$ and the fundamental cell volume $v_F$ are such that the relevant physical system (e.g. unjammed packings at some constant density or at some given distance to the jamming density $\phi_c$) is preserved.
In practice, we directly compute $\mc S(\vc k)$ from our simulation data and average the data from a large number of packings with the same system size to approximate $S(\vc k)$ for the ensemble of packings generated by a given protocol, keeping in mind that finite-system artifacts are expected to persist to some degree.
In all cases, the quantity $S(\vc 0)$ must be inferred through extrapolation to the origin.
Using this ensemble average, one may combine the Fourier transform of Eq.\ (\ref{eqn:OZ}) with Eq.\ (\ref{eqn:Sk}) in order to express the Fourier transform of the direct correlation function as
\begin{equation}
\tilde c(\vc k) = \frac{S(\vc k)-1}{\rho S(\vc k)},
\label{eqn:ck}
\end{equation}
from which $c(\vc r)$ may be computed by inverse Fourier transform.

A number of closure relations have been proposed to offer approximate solutions to Eq.\ (\ref{eqn:OZ}). 
In particular, the Percus-Yevick (PY) closure demands that $h(\vc r) = -1$ for $0 \le r < D$ and $c(\vc r)=0$ for $r>D$ for monodisperse spheres of unit diameter \footnote{To be mathematically precise, the statistical functions $g_2$, $h$, $S$, and $c$ are all functions which map $\reals^d \mapsto \reals$; however, since these functions are often isotropic, we often express them in terms of $r=\norm{\vc r}$ or $k=\norm{\vc k}$ as appropriate.}. 
Physically, these requirements specify that no two spheres may overlap, and that direct interactions (in the sense of Eq.\ (\ref{eqn:OZ})) are absent beyond the particles' hard cores. 
We would like the former criterion to apply in any solution for hard-spheres, but the latter assumption is increasingly violated as the packing fraction $\phi = \pi \rho/6$ increases. 
The cubic polynomial form of $c(\vc r)$ produced by the PY closure that solves Eq.\ (\ref{eqn:OZ}) accurately describes much of the equilibrium liquid branch of the hard-sphere system with good quantitative agreement for $\phi<0.40$ and qualitative agreement for $\phi<0.49$. 
However, it possesses various shortcomings at higher densities still within the liquid branch that one might like to improve: it underpredicts $g_2(1^+)$ (and thus the pressure), and oscillations in the pair correlation function are out of phase and decay too slowly with increasing $r$ \cite{Torquato_2002}. 
At higher densities, $g_2(r)$ fails basic satisfiability criteria such as nonnegativity and hence ceases to be physical.

In order to address this shortcoming, a variety of adjustments have been made to the PY approximation to improve the range of densities over which it may apply. 
A classic approach is to introduce a Yukawa term beyond the core \cite{Waisman_1973,Stell_1976_cr1,Stell_1976_cr2}, i.e., $c(r>D) = K e^{-z(r-D)}/r$. 
This improves the degree to which the approximation matches qualitative features in $c(\vc r)$ and provides additional fitting parameters to allow for a quantitative match of additional system properties. 
To this end, the recent work of Jadrich and Schweizer \cite{Schweizer_2013_II} used a two-Yukawa generalized mean spherical approximation which allowed them to match the system's compressibility as well as $g_2(D^+)$ and its first derivative in an attempt to accurately describe the behavior of the hard-sphere system along some metastable branch leading towards a disordered jammed state. 
By allowing $z$ and $K$ to approach infinity, this model may capture the appearance of a delta function at $c(r=D)$, and predicts a functional form for $c(\vc r)$ inside the core that departs from the solution to the Percus-Yevick approximation.
For a single-Yukawa form with $d=3$, one finds \cite{Waisman_1973}
\begin{eqnarray}
c(x) = &-&a - bx - \phi a x^3/2 
\nonumber
\\
 &-& \nu \frac{1-e^{-zx}}{zx} - \nu^2 \frac{\cosh (zx) - 1}{2 K z^2 e^z}
\label{eqn:WaismanYukawa}
\end{eqnarray}
for $x<1$, where $x=r/D$, $a=1-24\phi \int_0^\infty c(x)x^2 dx$, $\nu = 24 \phi \int_1^\infty x e^{-z(x-1)}g_2(x)dx$, $b$ satisfies
\begin{eqnarray}
24 \phi y_0^2 &=& -4b + 2 \nu z - \frac{\nu^2}{K e^z}
\\
24 \phi (y_1^2 - 2 y_0 y_2) &=& 24 \phi a - 2 \nu z^3 + \frac{\nu^2 z^2}{K e^z},
\end{eqnarray}
and $y_i = (d^i/dx^i)[x g_2(x)]_{D^+}$.
Interestingly, upon taking $k$ and $z$ towards infinity so as to obtain a delta function at $c(r=D)$, we see that the fourth term in Eq.\ (\ref{eqn:WaismanYukawa}) scales as $-1/r$ for $r_z<r$ as $r \rightarrow 0$, where $r_z \approx 1/z$.
Thus, as $z$ is taken to infinity, $r_z$ goes to zero.
For $r<r_z$, the term saturates to a constant.
Note that, in order for this scaling behavior to imply that $c(\vc r)$ diverges at the origin, one must have $\nu \rightarrow \infty$.
This is not necessarily implied by this analysis \cite{foot:Coulomb}.

While this approach improves upon these features, it still lacks the correct long-$r$ scaling behavior for MRJ-like packings, i.e., $c(\vc r) \propto -1/r^2$ as $r \rightarrow \infty$. 
Accordingly, the small-$k$ behavior of the structure factor is not in qualitative agreement either. 
In addition to this, several salient features in the pair correlation function, including (i) the power-law divergence as $r \rightarrow D^+$, (ii) the cusp at $r/D= \sqrt{3}$, and (iii) the correct step discontinuity at $r/D=2$, remain elusive.

\section{Inheritance of features in $c(\vc r)$ from $g_2(\vc r)$}
\label{sec:Inheritance}

Here, we present a derivation for the magnitude of the step discontinuity in $g_2(2)$ in terms of the information in $c(\vc r)$ for the specific case of disordered, jammed hard-sphere packings in three dimensions. 
Critically, we assume throughout this analysis that (i) $g_2$ and $c$ are isotropic (radial) functions and that (ii) the packings are isostatic, meaning that the average contact number for backbone spheres in the packing is $z=6+\order{1/N_B}$, where $N_B = (1-f_r) N$ is the number of backbone spheres in a packing with rattler fraction $f_r$. 
It is estimated that $f_r = 0.015$ for TJ and $f_r = 0.025$ for LS \cite{Atkinson_2013}. 
The particular form of the $\order{1/N_B}$ term depends upon whether one is considering collective or strict jamming, but becomes irrelevant in the infinite-system limit \cite{Donev_2005_PCF}. 
The pivotal observation from which this analysis follows is that a singularity of a given order in $c$ may not contribute to a singularity in $g_2$ of lower order.
For example, a step discontinuity in $c$ cannot cause a delta function to appear in $g_2$.
This assumption is justified by recursively inserting the form of $h(\vc r)$ into the Ornstein-Zernike relation to obtain
\begin{eqnarray}
h(\vc r_{12}) &=& c(\vc r_{12})
\nonumber 
\\
&~& + \rho \int_{\reals^3} c(\vc r_{13}) c(\vc r_{23}) d \vc r_3 
\nonumber
\\ 
&~& + \rho^2 \int_{\reals^3} c(\vc r_{13}) c(\vc r_{34}) c(\vc r_{24}) d \vc r_3 d \vc r_4
\nonumber
\\
&~& + \dots
\label{eqn:OZexpand}
\end{eqnarray}
and noting that each successive convolution ought to increase the order of the differentiability class of the term by one.

We begin with the delta function at $g_2(1)$.  By Eq.\ (\ref{eqn:OZexpand}), we see immediately that the only contribution to this must come from a delta function in $c(1)$, and that the two must be of equal magnitude.  The number of particles that are separated at a distance $r$ is given by $z(r) = \lim_{\epsilon \rightarrow 0} \int_{r-\epsilon}^{r+\epsilon} 4 \pi x^2 \rho g_2(x) dx$.  We substitute $z(1)=6f_b$, where $f_b=1-f_r$ is the backbone fraction, and obtain the result that the strength of the delta function in $g_2(1)$ is equal to $f_b/4\phi$.

We now proceed to identify the source of the jump discontinuity found at $g_2(r=2)$.  We begin by decomposing $c(r)$ into three parts:
\begin{equation}
c(\vc r) = c_\circ(\vc r) + c_\delta(\vc r) + c_\Theta(\vc r),
\label{eqn:crDecomposition}
\end{equation}
where $c_\delta(\vc r) = f_b \delta(r-1) / 4 \phi$ is the delta function contribution from above; $c_\Theta(\vc r) = A(1-\Theta(r-2))$, where $\Theta(x)$ is the Heaviside theta function, captures the step discontinuity predicted in $c$ at $r=2$, and $c_\circ(r)$ captures the remainder of the direct correlation and is assumed to be at least continuous at $r=2$.  Substituting this into Eq.\ (\ref{eqn:OZexpand}) gives
\begin{eqnarray}
h(\vc r_{12}) &=& c_\circ(\vc r_{12}) + c_\delta(\vc r_{12}) + c_\Theta(\vc r_{12})
\nonumber
\\
&~& + \rho \int_{\reals^3} \left[ \left(c_\circ(\vc r_{13}) + c_\delta(\vc r_{13}) + c_\Theta(\vc r_{13}) \right) \right.
\nonumber
\\
&~& \left. \left( c_\circ(\vc r_{23}) + c_\delta(\vc r_{23}) + c_\Theta(\vc r_{23}) \right) \right] d \vc r_3 + \dots
\end{eqnarray}
Substituting in $r_{12}=2+\epsilon$ and $r_{12}=2-\epsilon$ into this and subtracting the two equations from each other while letting $\epsilon \rightarrow 0$ gives:
\begin{equation}
  \Delta_\epsilon h(2) = \Delta_\epsilon c_{\circ }(2) + \Delta_\epsilon c_{\delta}(2) + \Delta_\epsilon c_{\Theta}(2) + \rho \Delta_\epsilon I(2),
\end{equation}
where we have used the shorthand $\Delta_\epsilon f(x) = \lim_{\epsilon \rightarrow 0} f(x+\epsilon) - f(x-\epsilon)$ and defined
\begin{equation}
I(\vc r_{12}) = \int_{\reals^3} c(\vc r_{13}) c(\vc r_{23}) d \vc r_3.
\label{eqn:crIntegralTerm}
\end{equation}
We notice that $\Delta_\epsilon c_\delta(2) = \Delta_\epsilon c_\circ(2) = 0$, meaning that the only surviving term of the first three is $\Delta_\epsilon c_\Theta(2) = -A$.
Furthermore, by considering the decomposition of Eq.\ (\ref{eqn:crDecomposition}) as applied to Eq.\ (\ref{eqn:crIntegralTerm}), we can see that the term with the differentiability class of the lowest degree is given by
\begin{equation}
I_\delta(\vc r_{12}) = \int_{\reals^3} c_\delta(\vc r_{13}) c_\delta(\vc r_{23}) d \vc r_3,
\end{equation}
and that $\Delta_\epsilon I_\delta(2)$ should be nonzero so long as the delta function represented by $c_\delta$ has a nonzero amplitude (i.e.\ $c_\delta$ is not trivially zero everywhere).  All other terms contributing to $I(r)$ are continuous at $r=2$, as are all higher-order convolutions. 
Thus, we are left with
\begin{equation}
  \Delta_\epsilon h(2) = \rho \Delta_\epsilon I_\delta(2) - A.
\end{equation}
To evaluate the first term, we define $\bar{\vc r} = (\vc r_1 + \vc r_2)/2$, so that
\begin{equation}
  I_\delta(\vc r_{12}) = 
  \int_{\reals^3} 
  c_\delta \left(\vc r_3 - \bar{\vc r} + \frac{\vc r_{12}}{2} \right) 
  c_\delta \left(\vc r_3 - \bar{\vc r} - \frac{\vc r_{12}}{2} \right) 
  d\vc r_3.
\end{equation}
We invoke the convolution theorem to write 
\begin{equation}
  \tilde I_\delta(k) = \tilde c_\delta(k)^2,
\nonumber
\end{equation}
where we have invoked translational invariance to remove $\bar{\vc r}$ and the symmetry of the two terms within the integrand with respect to $\vc r_{12}$.
Carrying out the inverse Fourier transform gives the result
\begin{eqnarray}
  \rho I_\delta(\vc r_{12}) &=& \frac{\rho}{2 \pi^2 r} \int_0^\infty \tilde c_\delta^2(\vc k) k \sin kr dk
  \nonumber
  \\
  &=& \frac{\rho}{2 \pi^2 r} \int_0^\infty \left(\frac{\pi}{\phi k} \sin k \right)^2 k \sin kr dk
  \nonumber
  \\
  &=& \frac{3}{\pi \phi r} \int_0^\infty \frac{1}{k} \sin^2 k \sin kr dk
  \nonumber
  \\
  &=& \frac{3}{4 \phi r} \left( 1 - \Theta(r-2) \right).
\end{eqnarray}
Because $I_\delta$ represents the ``sharpest'' contribution from the single convolution term, we can conclude immediately that no other terms within the first convolution term (and no further convolution terms) will contribute to the quantity $\lim_{\epsilon \rightarrow 0} h(2+\epsilon) - h(2-\epsilon)$ since they will be too smooth (i.e.\ they do not retain a step discontinuity following convolution).
Thus, we arrive at the result that the magnitude of the step discontinuity in the total correlation function (and, equivalently, the pair correlation function) at $r=2$ is
\begin{equation}
  \Delta_\epsilon h(2) = \Delta_\epsilon c(2) - \frac{3}{8 \phi}.
  \label{eqn:Step}
\end{equation}
By extending this analysis, we can claim that {\it even in the absence of any additional nonanalytic features in $c(\vc r)$,} $g_2(\vc r)$ is expected to have discontinuities in successively higher derivatives at further integer values of $r$.
For example, there ought to be a discontinuity in the first derivative of $g_2$ at $r=3D$, a discontinuity in the second derivative at $r=4D$, and so on.

\section{Scaling relations for systems in the vicinity of hyperuniformity}
\label{sec:Scaling}

In this section, we recall various scaling behaviors for various pair statistics in direct and Fourier spaces for ordered and disordered packings of hard spheres in the vicinity of jamming that are particularly germane to this paper.

Torquato and Stillinger have shown \cite{Torquato_2003_Hyperuniform} that a hyperuniform system with a structure factor that scales as
\begin{equation}
S(\vc k) \propto k^{2-\eta}, ~~k \rightarrow 0
\label{eqn:skscaling}
\end{equation}
may be thought of as an ``inverted critical point.''
At this point, the direct correlation becomes long-ranged, scaling as
\begin{equation}
c(\vc r) \propto -r^{2-d-\eta}, ~~r \rightarrow \infty
\label{eqn:crscaling}
\end{equation}
in dimension $d$, where $\eta$ is a critical exponent such that $2-d < \eta < 2$. 
Additionally, the inverse of the structure factor at the origin exhibits critical scaling behavior in the vicinity of its critical density, i.e., 
\begin{equation}
S^{-1}(\vc 0) \propto (1-\phi/\phi_c)^{-\gamma}
\label{eqn:HUexpgamma}
\end{equation}
for densities close to, but below $\phi_c$ \cite{Torquato_2003_Hyperuniform}.

In the case of the equilibrium crystal branch, we may exploit the compressibility relation relating the structure factor to the isothermal compressibility $\kappa_T$ at temperature $T$
\begin{equation}
S(\vc 0)=\rho k_B T \kappa_T
\label{eqn:compressibility}
\end{equation}
along with the free-volume equation of state \cite{Salsburg_1962} which predicts that the pressure $p$ behaves as
\begin{equation}
\frac{p}{\rho k_BT} = \frac{d}{1-\phi/\phi_J},
\label{eqn:FVEOS}
\end{equation}
to obtain the result that $\kappa_T \propto (1 - \phi/\phi_c)^{2}$ in the vicinity of jamming.
From this, it follows that
\begin{equation}
\gamma = 2 ~\text{(equilibrium crystal)}.
\label{eqn:gammaFCC}
\end{equation}
Note that this result is independent of dimension.
One may also define a correlation length $\xi$ with the critical behavior
\begin{equation}
\xi \propto (1-\phi/\phi_c)^{-\nu}
\end{equation}
which may be related to the previous critical exponents through
\begin{equation}
\gamma = (2-\eta)\nu.
\label{eqn:exponents}
\end{equation}
As mentioned before, one such length scale can be defined by the volume integral of the direct correlation function: $\xi = (-\int_{\reals^3} c(\vc r) d \vc r)^{1/d}$.

For packings along glassy metastable branches leading to MRJ-like states, one cannot use the compressibility relation because the states are nonequilibrium in nature \cite{Hopkins_2012}.  However, one may reconcile the differing pictures presented by $S(\vc 0)$ and $\kappa_T$ by introducing a ``nonequilibrium index'', defined as
\begin{equation}
X \equiv \frac{S(\vc 0)}{\rho k_BT \kappa_T} - 1.
\label{eqn:nonequilibriumindex}
\end{equation}
Hopkins et al.\ studied the behavior of $X$ under rapid compression toward jamming for MRJ-like packings prepared by the LS algorithm for system sizes up to $N=10^6$ and found that, as $\phi$ increased toward jamming, $X \propto (1-\phi/\phi_c)^{-1}$ \cite{Hopkins_2012}.

By combining this result with the observation that the pressure in MRJ-like packings again diverges according to the free-volume equation of state, we get the result that 
\begin{equation}
\gamma = 1 ~\text{(MRJ packings)}.
\end{equation}
By Eq.\ (\ref{eqn:exponents}), this implies that $\nu=1$ for MRJ packings as well.
This is a noteworthy result because it tells us that while both MRJ-like and equilibrium crystalline packings become increasingly hyperuniform as they are compressed, {\it they are different universality classes with respect to the critical exponents $\eta$ and $\gamma$ associated with hyperuniformity.}
While ensembles of packings are known to exist that interpolate between these extremes, the interpolation is not unique, owing to the large diversity of jammed packings that are known to exist.
Thus, it is an interesting, outstanding question how these critical exponents might evolve between these two extreme states.

%\red{Critical exponent $\beta$ that, in percolation theory, describes the critical scaling of $P(p-p_c)$ (probability that a given site/bond is within the percolating cluster ).  What is this here?}

%%%%%%%%%%%%%%%%%%%%%%%%%%%%%%%%%%%
%%            METHOD             %%
%%%%%%%%%%%%%%%%%%%%%%%%%%%%%%%%%%%

\section{Simulation Methods}
\label{sec:Methods}

In order to study the behavior of the equilibrium hard-sphere FCC crystal for densities between $\phi=0.55$ and the close-packing density $\phi_{FCC} = \pi / \sqrt{18}$, we used standard event-driven molecular dynamics \cite{Skoge_2006}. 
Configurations of $N=4M^3$ spheres with $4 \le M \le 63$ were placed on their lattice sites and allowed to equilibrate at fixed packing fraction within a cubic fundamental cell with periodic boundary conditions for $10^5$ collisions per sphere before taking statistics. 
Measurements of the structure factor were made every $10^3$ collisions per sphere to verify that equilibrium had been attained.

To generate disordered, strictly jammed sphere packings in three dimensions, we begin with initial conditions produced by random sequential addition (RSA) at an initial packing fraction of $\phi=0.10$. 
We use the Torquato-Jiao (TJ) sequential linear programming method \cite{TJ_2010} with system sizes of up to $N=10^4$ using the same parameters as those used to study the MRJ state in Ref.\ \cite{Atkinson_2013}. 
The final mean density of these packings is $\phi=0.6352 \pm 2.6 \times 10^{-4}$ for a system size of $N=2000$. 
After the algorithm terminates and a putatively jammed state is reached, the packing is equilibrated within its jamming basin using event-driven molecular dynamics at fixed density. 
We also compare these packings to those generated using the well-known event-driven Lubachevsky-Stillinger (LS) molecular dynamics algorithm \cite{Stillinger_1990}. 
For LS, we use an initial dimensionless growth rate of $\Gamma = dD/dt \sqrt{m/(k_BT)} = 10^{-2}$ until packings reach a dimensionless pressure of $P = pV/Nk_B T = 10^4$, at which point the expansion rate is slowed to $\gamma=10^{-5}$, and packing continues until $P=10^8$. 
The mean density of the final packings as prepared under this protocol is $\phi = 0.6434 \pm 1.0 \times 10^{-4}$ at a system size of $N=10000$.

%%%%%%%%%%%%%%%%%%%%%%%%%%%%%%%%%%%
%%            RESULTS            %%
%%%%%%%%%%%%%%%%%%%%%%%%%%%%%%%%%%%
%
% Be sure to include:
%
% delta function doesn't explain the step at x=2
%
%
\section{Results}
\label{sec:Results}

In this section, we present results pertaining to our computer-generated hard-sphere packings as they approach both ordered and disordered jammed states.  We will present our analysis assuming that the particle diameter $D$ is unity unless otherwise specified.

%=====================================================================================
% FCC

\subsection{FCC}
We begin by examining the behavior of the hard-sphere FCC crystal because the behavior exhibited by the crystal undergoing thermal motion away from jamming serves as a interesting starting point from which we can make several observations to guide our subsequent investigation of disordered jammed systems.
We will investigate the crystal's approach to close-packing with attention to the fluctuating component of the structure factor as well as the implications it has for the qualitative form of $c(\vc r)$.

Figure \ref{fig:sk_FCC} shows plots of the radially-averaged structure factor $S(\vc k)$ for the equilibrated FCC crystal for a variety of densities along the solid branch; our computation follows the collective coordinate formulation of Eq.\ (\ref{eqn:SkCC}).
Curves are averaged over ensembles of 100 packings with $N=2048$. 
Figure \ref{fig:cr_FCC} shows the corresponding radially-averaged direct correlation function evaluated numerically using discrete Fourier transform techniques following Eq.\ (\ref{eqn:ck}).

% S(k) of LS-equilibrated FCC packings for N=2048
\begin{figure}[hbt]
  \begin{centering}
    \includegraphics[width=0.5\textwidth,clip]{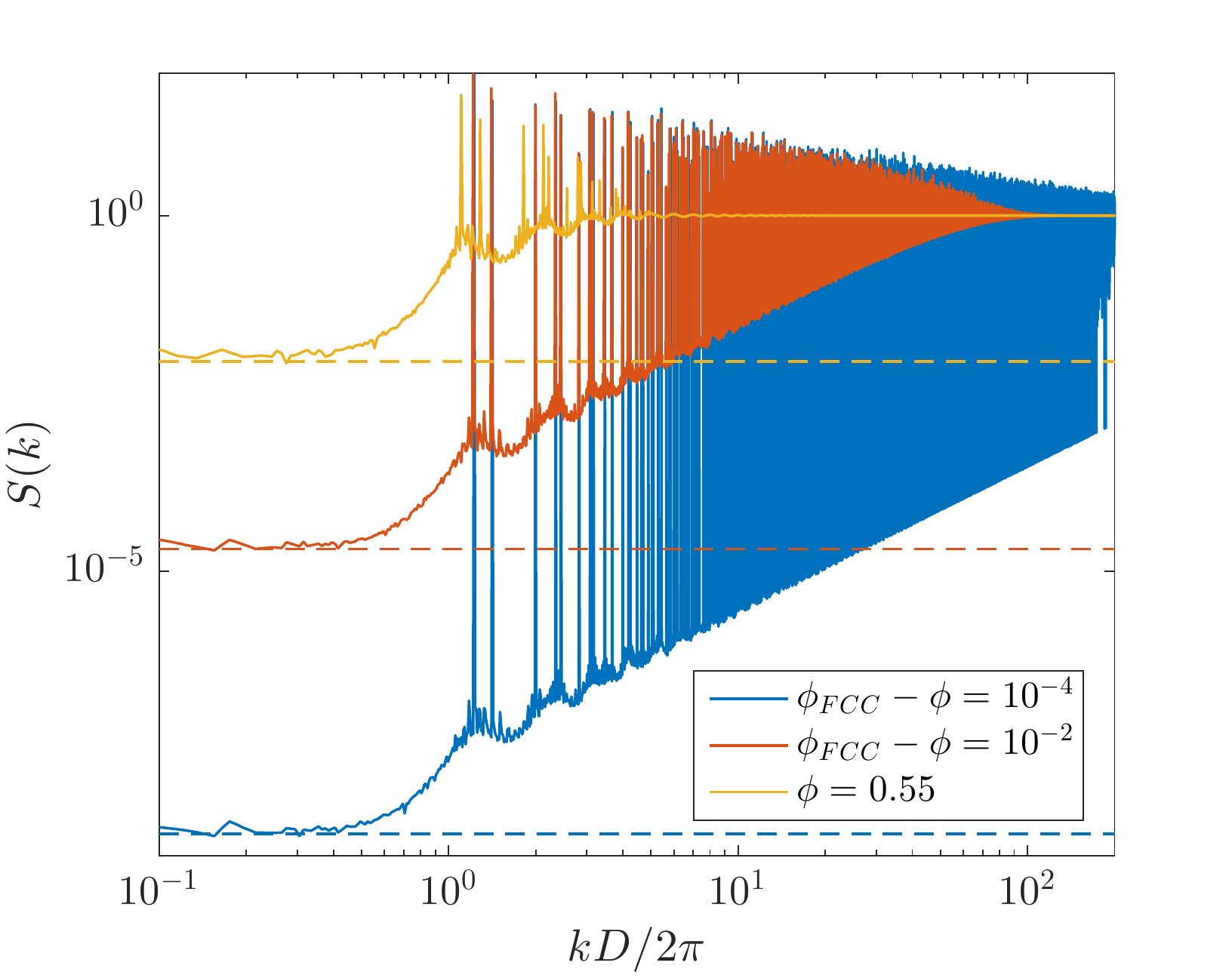}
    \caption{(Color online.) The structure factor of the hard-sphere FCC crystal at various densities and system size $N=2048$.  Dashed lines showing $S=(1-(\phi/\phi_{FCC})^{1/d})^2$ highlighting the critical behavior of $S(\vc 0)$ are included.}
    \label{fig:sk_FCC}
  \end{centering}
\end{figure}

% c(r) is for the same data
\begin{figure}[hbt]
  \begin{centering}
    \includegraphics[width=0.5\textwidth,clip]{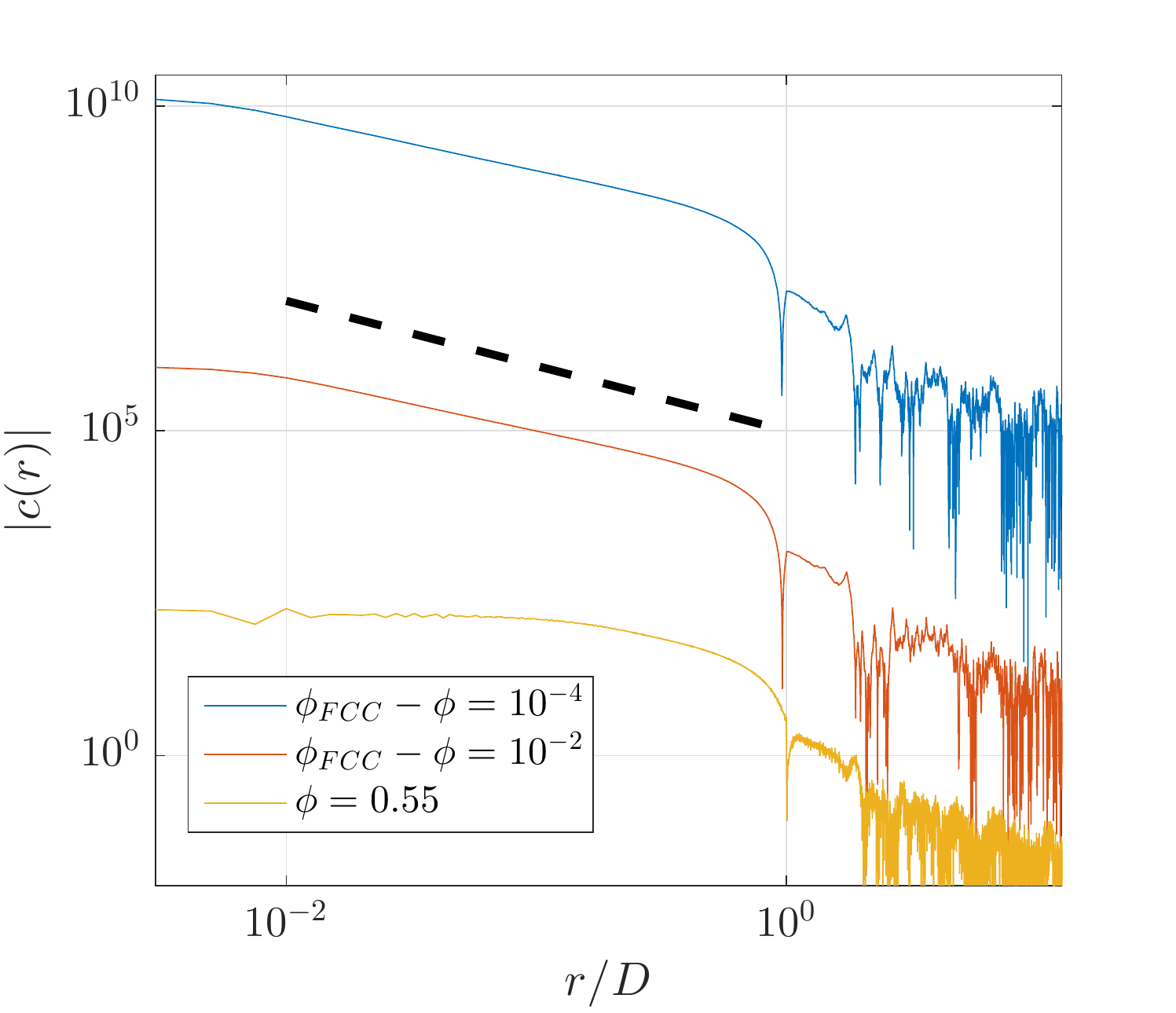}
    \caption{(Color online.) The radially-averaged direct correlation function of the hard-sphere FCC crystal at various densities and system size $N=2048$.  The thick dashed line is included as a guide for the eye and shows a $1/r$ scaling as $r \rightarrow 0$.}
    \label{fig:cr_FCC}
  \end{centering}
\end{figure}

In addition to the Bragg peaks that arise from the crystal geometry from the ``frozen-in'' structure of the packings, the curves possess a background contribution derived from thermal fluctuations which scales as $k^{2}$ starting at a sufficiently-high wavenumber (on the order of unity) and saturates to $S=1$. 
While the domains of interest in the scaling descriptions of Eqs.\ (\ref{eqn:skscaling}) and (\ref{eqn:crscaling}) as written are $k \rightarrow 0$ and $r \rightarrow \infty$, one might reasonably suspect that it is possible to invert the two limits so as to infer the scaling behavior of $c(r \rightarrow 0)$ from $S(k \rightarrow \infty)$. 
In the case of our FCC data, we see that this implies that the direct correlation function ought to scale as $r^{-1}$ for $r<1$. 
Looking at Fig.\ \ref{fig:cr_FCC}, this seems to be the case. 
We note in passing that this is also consistent with the prediction given by the analysis of the Yukawa form mentioned in Sec.\ \ref{sec:Scaling}.

At sufficiently low wavenumbers, $S(\vc k)$ converges to a constant, given to a good approximation as $S(\vc 0) = (1-(\phi/\phi_{FCC})^{1/3})^2$, confirming that the critical exponent $\gamma = 2$ as predicted in Eq.\ (\ref{eqn:gammaFCC}).
Importantly, at the point of exact jamming, the FCC crystal is trivially stealthy (its structure factor is identically zero up to some positive wavenumber), meaning that the critical exponent $\eta$ may be thought of as being infinite.
This discontinuous change from the equilibrium behavior at even vanishingly small distances to jamming highlights the singular nature of jamming and underscores the need to be careful when considering limiting behavior.

It is important to note that a structure factor that scales as $k^2$ can be obtained by applying random, uncorrelated displacements to each particle in the crystal \cite{Gabrielli_2004}. 
Given the inherent anharmonicity of the system owing to the singular nature of the hard-sphere interaction potential, we are motivated to ask whether the probability distribution of the pair separation of nearest-neighbor spheres in the crystal might be, to a good approximation, statistically independent from pair to pair on certain length scales.
On larger length scales (lower wavenumbers), the exact form of $S(\vc k)$ is in excellent agreement with a normal mode analysis based on the fictitious interparticle potential derived in \cite{Brito_2006}, suggesting underlying correlated displacements on corresponding length scales.
In the second paper in this series, we investigate the consequences of this by analyzing the percolation properties of an intimately-related ``cherrypit model'' \cite{Atkinson_2016_Perc}.

%=====================================================================================
% Disordered
\subsection{Disordered Packings and the MRJ state}

We now turn our attention to characterizing the approach to jamming in disordered packings produced by the LS and TJ algorithms as described in Sec.\ \ref{sec:Methods}.
Unlike the case for the FCC crystal, MRJ-like packings do not strongly indicate signs of an incipient jammed structure at densities far below that of jamming.  Furthermore, there is a strong protocol dependence yielding {\it qualitative} differences in various packing protocols' approaches towards disordered, jammed states.

%Before jamming
Figure \ref{fig:SkDisordered} shows the ensemble-averaged structure factor of our packings for a variety of densities and protocols. 
As jamming is approached, $\lim_{k \rightarrow 0} S(\vc k)$ approaches zero, implying that the jammed state is hyperuniform, in agreement with previous investigations \cite{Donev_2005_Unexpected,Hopkins_2012}.
Because the data is presented on a log-log scale, the vertical offset between the nearly-jammed TJ and LS configurations corresponds to a difference in slope in their respective linear behavior in the vicinity of the origin.
Interestingly, the packings produced by the TJ algorithm display anomalous behavior well before jamming is reached, including a structure factor that {\it increases} as the wavenumber approaches zero for intermediate densities.
By contrast, $S(\vc k)$ for LS-generated packings seems to monotonically decrease for $kD/2\pi < 1$ at all packing fractions leading up to jamming. 
This suggests that the intermediate configurations that TJ creates on its way to jamming are far from equilibrium---even at packing fractions below the freezing density $\phi_f \approx 0.494$ \cite{Hoover_1968}. 
For configurations that are close to jamming at a density of $\phi_c$, we group packings according to the quantity $1-\phi/\phi_c$ rather than $\phi$.

\begin{figure}[hbt]
  \begin{centering}
    \includegraphics[width=0.5\textwidth,clip]{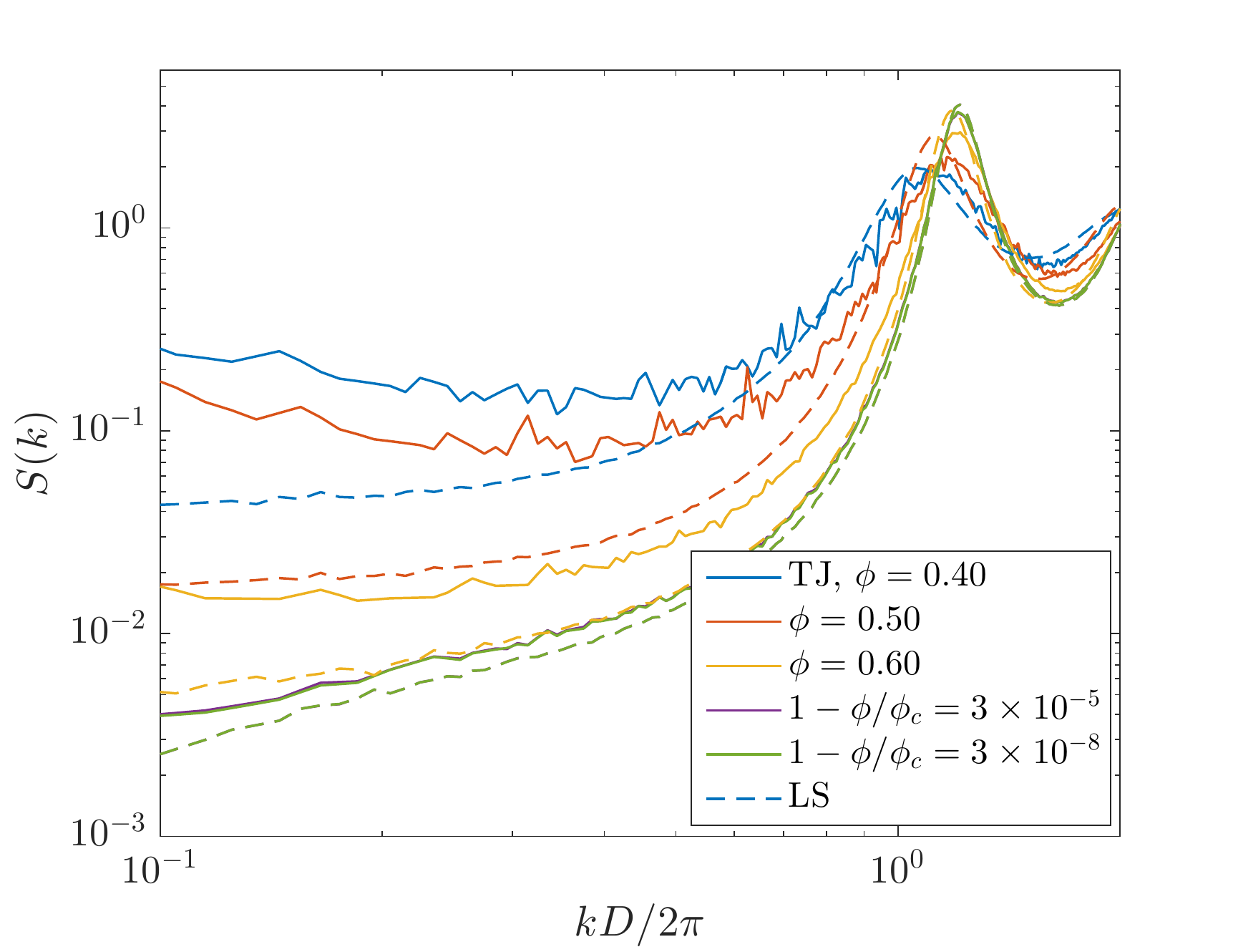}
    \caption{(Color online.) The structure factor for ensembles of packings created using TJ (solid lines) and LS (dashed lines) at various packing fractions $\phi$ and distances to jamming $1-\phi/\phi_c$.  Note that very little change is observed upon compressing the systems from $1-\phi/\phi_c = 3 \times 10^{-5}$ to $3 \times 10^{-8}$.}
    \label{fig:SkDisordered}
  \end{centering}
\end{figure}

The TJ algorithm seems to find configurations that are consistently more disordered than those visited by LS starting at intermediate densities and continuing up to jamming \footnote{for very low densities, the RSA configurations that are given to TJ as initial conditions are rather similar to the equilibrium fluid.}.
One may consider the order metric
\begin{equation}
\tau = \frac{1}{(2\pi)^d} \int_{\norm{\vc k} < k_{max}} (S(\vc k)-1)^2 d \vc k,
\label{eqn:tau}
\end{equation}
which may be thought of as quantifying extent to which a given configuration differs from a Poisson process (for which $S=1$ for all $k$).
This order metric was used in \cite{Zhang_2015} with $k_{max} \rightarrow \infty$.
The order metric $\tau$ is also reminiscent of the direct-space order metric $T^*$ that we have utilized before \cite{Atkinson_2013}, which measures deviations in the pair correlation function from unity, as well as the two-body excess entropy $s^{(2)}$ \cite{Truskett_2000,Green_1952,Nettleton_1958,Baranyai_1989}.
Here, we keep $k_{max}$ finite to prevent $\tau$ from diverging due to the contributions of Dirac delta functions in the corresponding direct-space statistics characteristic of jammed packings; a similar issue arises in the aforementioned order metrics as discussed in \cite{Truskett_2000}.

Looking at Fig.\ \ref{fig:SkDisordered}, one can see immediately that $S(\vc k)$ for TJ-generated packings remains much closer to unity at intermediate densities than for LS.
This difference persists up to jamming.
Figure \ref{fig:tau} shows the quantity $\tau$ computed for ensembles of 1000 packings created by the TJ and LS algorithms corresponding to the densities used for Fig.\ \ref{fig:SkDisordered} as a function of $k_{max}$.
Consistent with other order metrics including $T^*$ as well as the standard bond-orientational order metric $Q_6$, which primarily measure short-range order, $\tau$ demonstrates that the packings generated by TJ are also more disordered than those produced by LS on larger length scales \cite{Atkinson_2013}.
Thus, $\tau$ provides complementary information to $T^*$ and $Q_6$.

\begin{figure}[hbt]
  \begin{centering}
    \includegraphics[width=0.5\textwidth,clip]{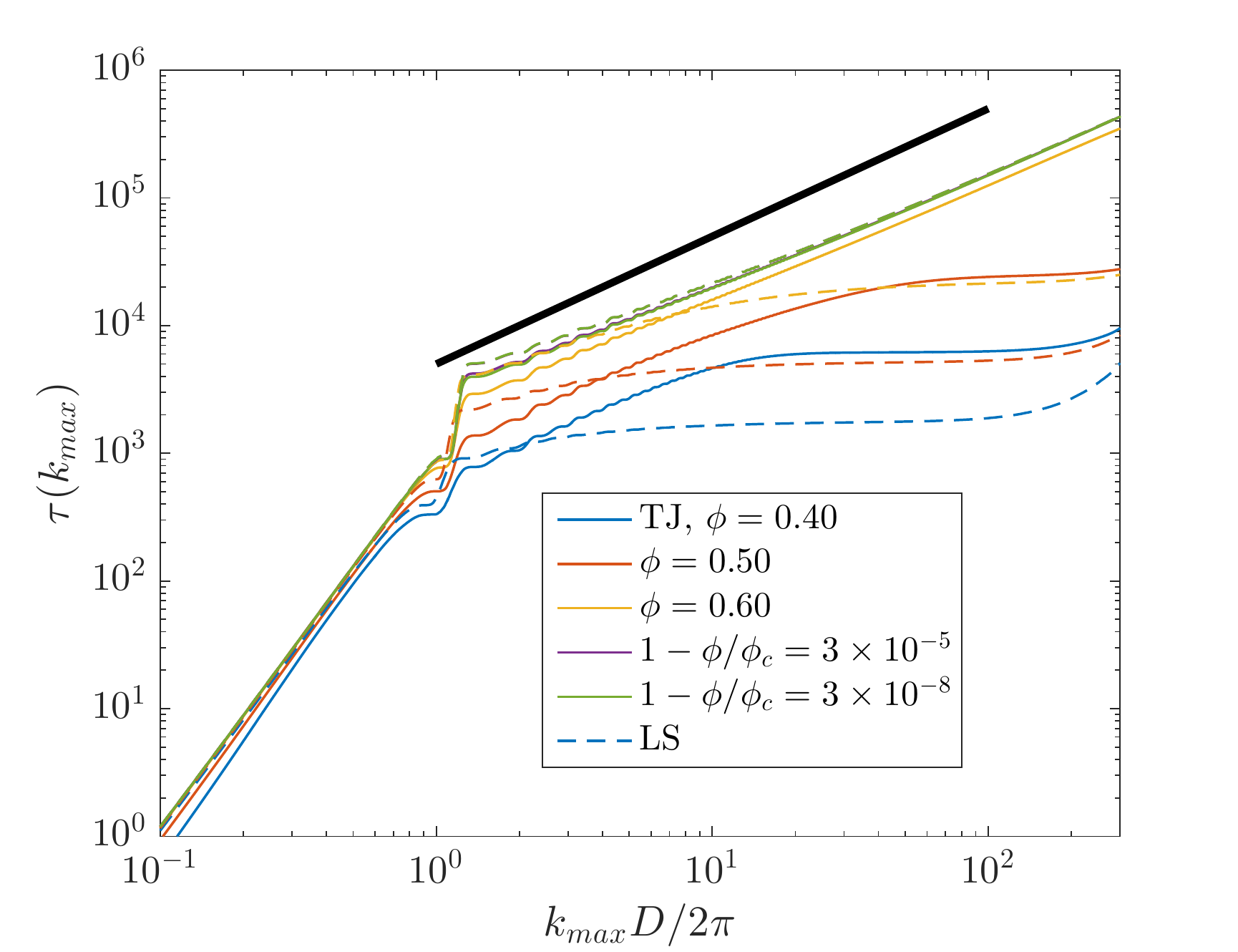}    
    \caption{(Color online.) The order metric $\tau$ for ensembles of packings created using TJ (solid lines) and LS (dashed lines) at various packing fractions $\phi$ and distances to jamming $1-\phi/\phi_c$ for different cutoffs $k_{max}$. Note that very little change is observed upon compressing the systems from $1-\phi/\phi_c = 3 \times 10^{-5}$ to $3 \times 10^{-8}$.}
    \label{fig:tau}
  \end{centering}
\end{figure}

Near jamming and at sufficiently high wavenumbers, the behavior of the integrand in Eq.\ (\ref{eqn:tau}) is dominated by the contribution arising from the delta function in $c$ at $r=D$.
This causes $\tau$ to diverge toward infinity at a rate that is linear in $k_{max}$.
The thick black line in Fig.\ \ref{fig:tau} illustrates the slope associated with this behavior.
If the delta function is not sharp (because of a spread in nearest-neighbor distances), then $\tau$ will saturate to a constant.
This is seen for the LS and TJ packings for $\phi=0.40$ and $0.50$ as well as for LS at $\phi=0.60$.
This suggests that, while near-contacts accumulate starting at low densities as TJ densifies its packings, they remain spread out over a range of pair distances beyond contact.

Because of the noise in measuring the structure factor numerically (due to both a finite number of packings in the ensemble as well as finite system sizes), $S(\vc k)$ the decaying oscillations converging to $S=1$ will eventually saturate to white noise.
Therefore, $\tau$ will begin to grow with increasing cutoff as $k_{max}^d$.
The beginning of this trend is visible in Fig.\ \ref{fig:tau} for our ensembles at lower densities.

Figure \ref{fig:crDisordered} shows the corresponding direct correlation functions for these ensembles of packings. 
The first salient feature is that the direct correlation function for the packings produced by the TJ algorithm exhibits a prominent peak at $r=1$ that is clearly visible at packing fractions as low as $\phi=0.50$. 
This is accompanied by a steep decrease in $c(\vc r)$ for $r<1$ that is dominated by a $-1/r$ scaling as $r \rightarrow 0$.
As mentioned in Sec.\ \ref{sec:OZJamming}, the analysis of a Yukawa-like $c(\vc r)$ beyond the core as $z$ goes to infinity \cite{Waisman_1973,Henderson_1975,Stell_1976_cr1,Stell_1976_cr2,Schweizer_2013_II} provides the rationalization for the appearance of a scaling behavior of this form.  However, it only becomes dominant if the volume integral of $c(\vc r)$ outside the core is sufficiently large.
That is, the short-ranged behavior of $c(\vc r)$ seems to be, fascinatingly, communicating its growing long-rangedness in the sense of the theoretical considerations of Sec.\ \ref{sec:Scaling}.

The early appearance of a delta function in $c$ at $r=1$ leads us to suggest that cluster formation in TJ occurs long before the packing is confined to a jamming basin. 
This result is likely related to the observation of Shen et al.\ \cite{OHern_2012} that athermal packings of spheres compressed from low densities in the presence of a viscous background exhibit a ``contact percolation'' which is accompanied by the emergence of a nontrivial mechanical response to applied stress at densities significantly below that of jamming. 
Because of the manner in which the sequential linear programming algorithm searches for {\it local} optimizations in packing fraction, which require little reconfiguration at low densities, there is reason to believe that the TJ algorithm explores available configuration space in a similar fashion to the procedure of Shen et al.\ for low to intermediate densities.

\begin{figure*}[hbt!]
  \begin{centering}
    $\begin{array}{cc}
    \includegraphics[width=0.5\textwidth,clip]{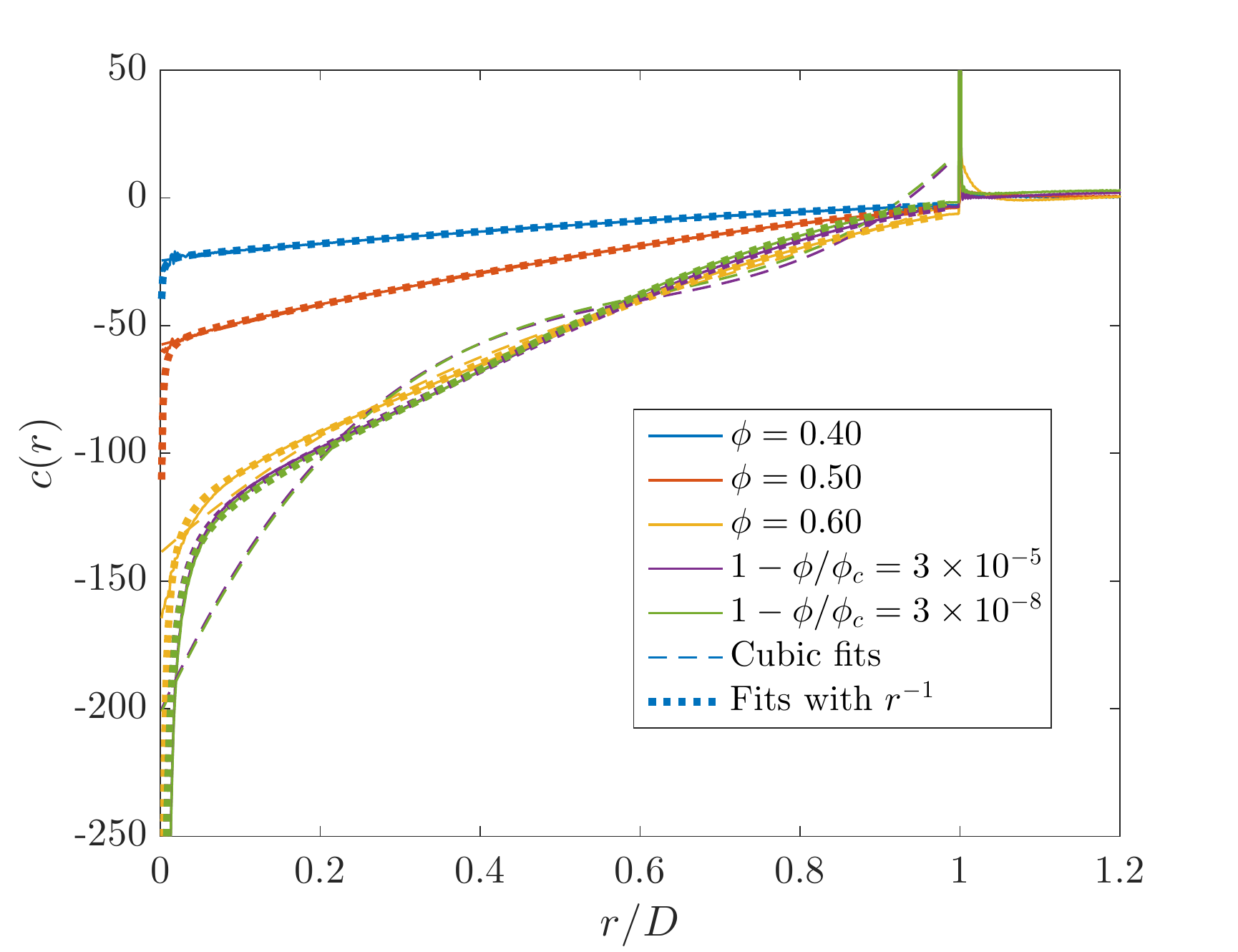}
      &
    \includegraphics[width=0.5\textwidth,clip]{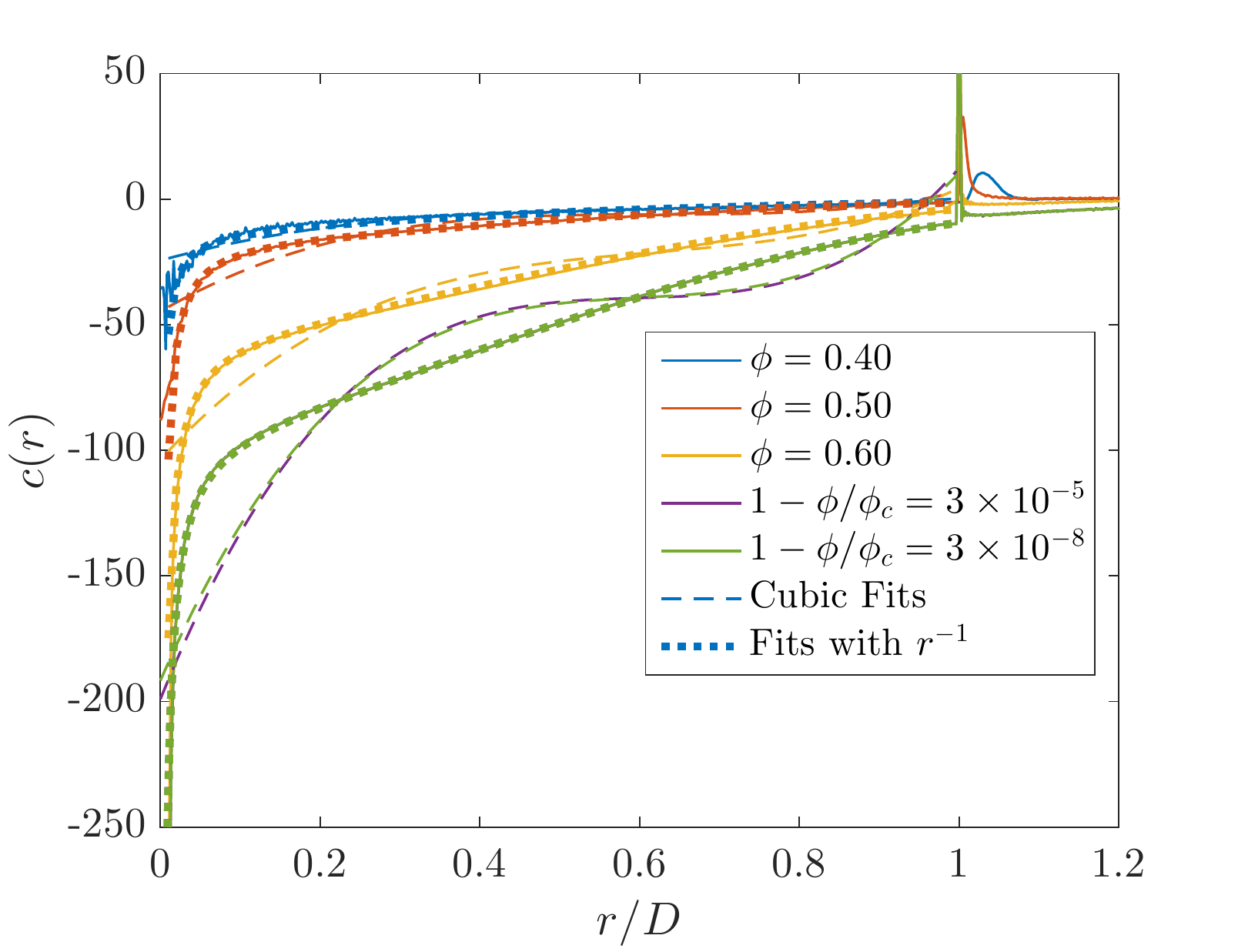}
      \\
      \mbox{(a)}
      &
      \mbox{(b)}
    \end{array}$
    \caption{(Color online.) Direct correlation functions computed for packings generated by (a) LS and (b) TJ for various densities.}
    \label{fig:crDisordered}
  \end{centering}
\end{figure*}

We also note in passing that the direct correlation functions exhibit both a cusp at $r=\sqrt{3}$ and a mild step discontinuity at $r=2$, mirroring the features found in the pair correlation function. 
We have noticed that the step discontinuity observed in $g_2$ of jammed packings is consistently larger than what is produced as an effect of the delta function at $c(r=1)$; this is explained in light of the result contained in Eq.\ (\ref{eqn:Step}).
Note also that the particular form of Eq.\ (\ref{eqn:Step}) relies on the assumption that the packing is isostatic, meaning that there are the minimum number of backbone contact pairs necessary to ensure jamming. 
In the infinite system limit, this corresponds to a mean backbone coordination number of $z=6 + \order{1/N}$, where the vanishing term reflects the difference between collective or strict jamming \cite{Donev_2005_PCF}.

Figure \ref{fig:crDisorderedLog} shows $-c(\vc r)$ plotted on a log scale. 
As jamming is approached, we observe that $c \propto -1/r^2$ for large $r$, confirming numerically the prediction of Eq.\ (\ref{eqn:crscaling}) for $\eta=1$ in the case of disordered packings
This scaling behavior is difficult to obtain numerically since one must accurately obtain $S(\vc k)$ data for low wavenumbers in order to extract the large-$r$ behavior of $c$. 
In particular, one must necessarily extrapolate $S(0)$, as a direct computation using Eq.\ (\ref{eqn:SkCC}) contains a forward scattering contribution which must be omitted. 
Additionally, to improve numerical stability, our $\tilde c(\vc k)$ data were multiplied by the Fourier transform of a narrow triangular window so that the real-space data is smoothed accordingly via convolution. 
Details of the procedure are give in Appendix \ref{sec:appendix_crNumerics}.

\begin{figure*}[hbt!]
  \begin{centering}
    $\begin{array}{cc}
    \includegraphics[width=0.5\textwidth,clip]{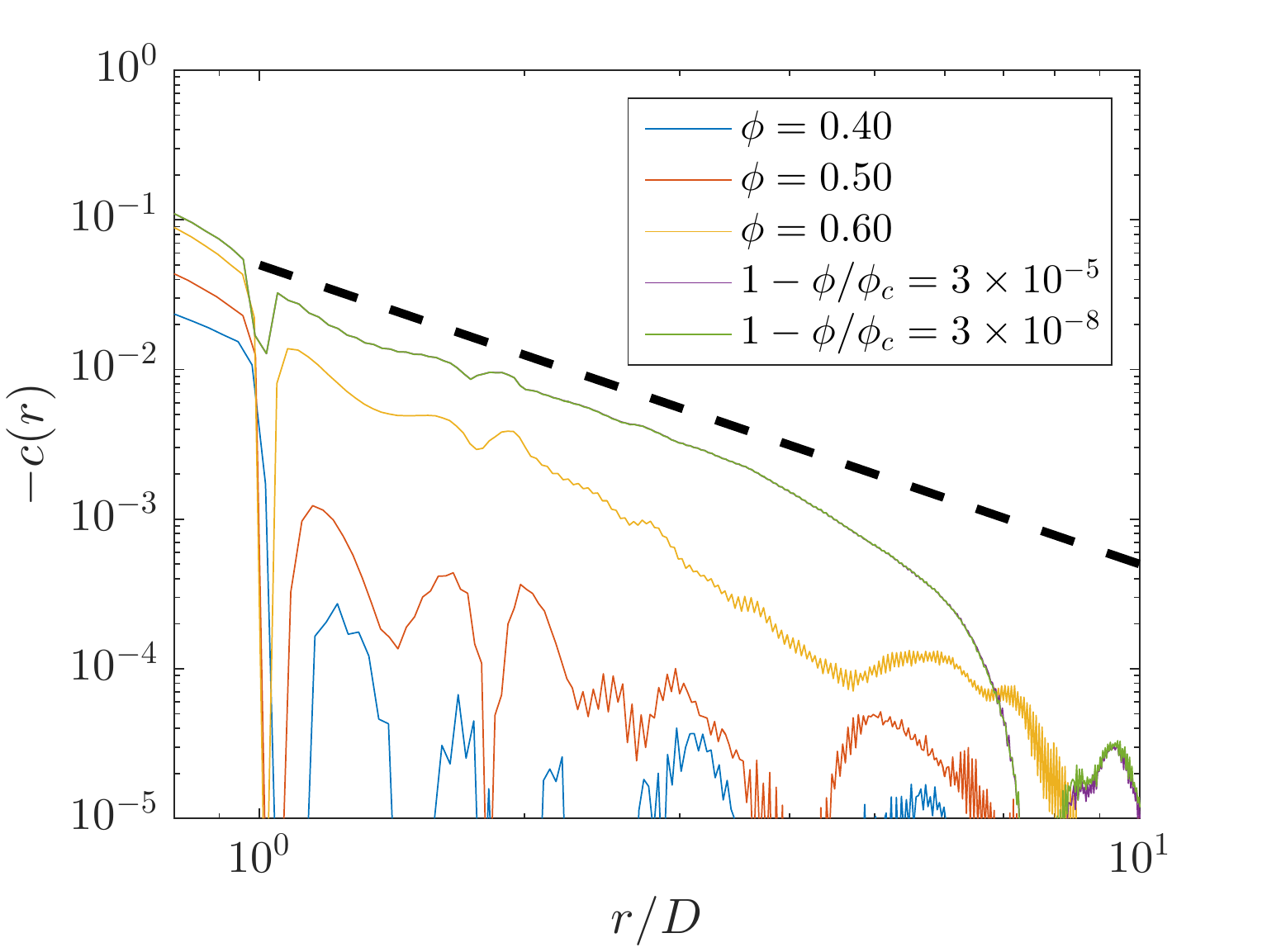}
      &
    \includegraphics[width=0.5\textwidth,clip]{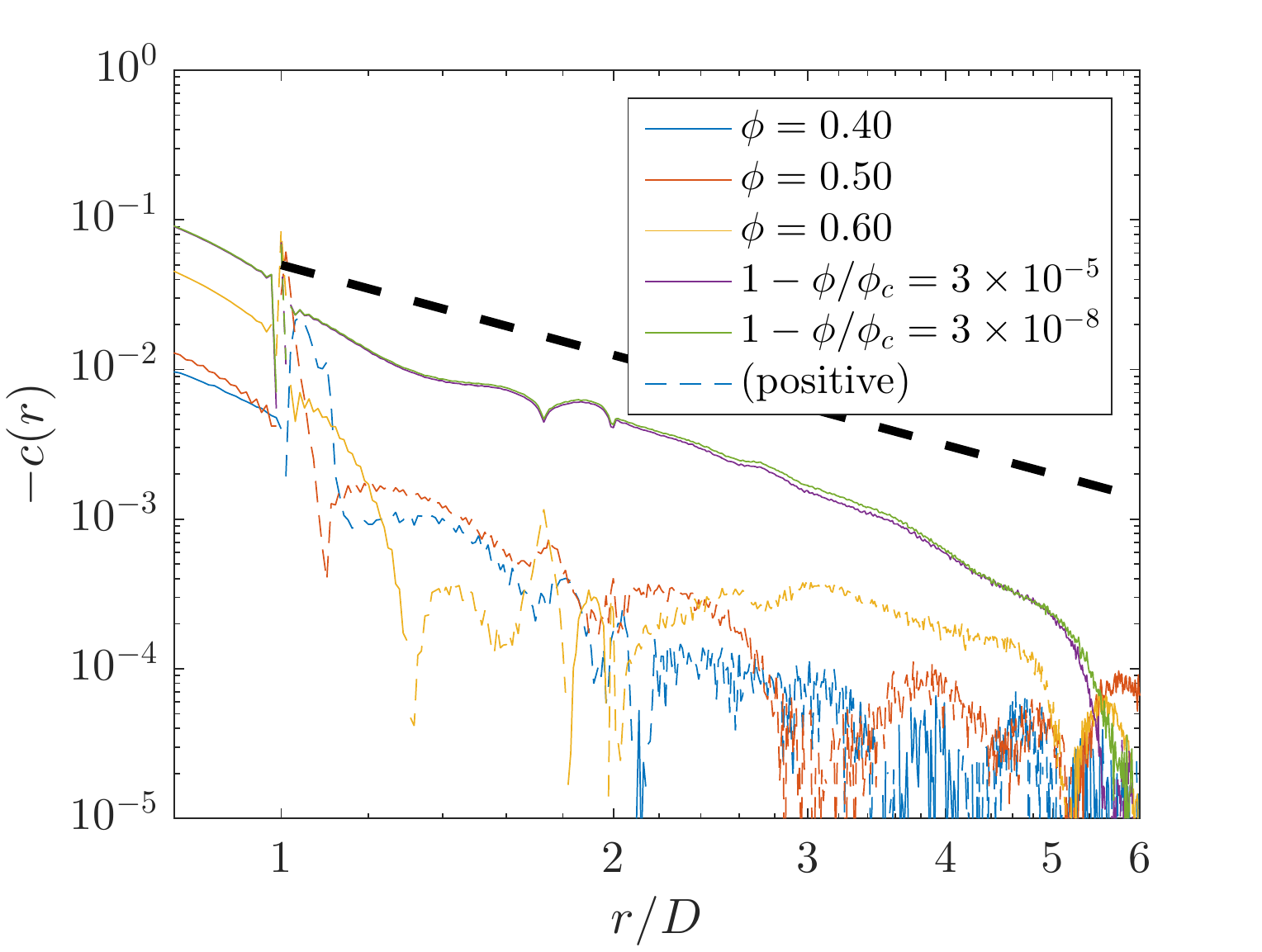}
      \\
      \mbox{(a)}
      &
      \mbox{(b)}
    \end{array}$
    \caption{(Color online.) Log-log representation of $-c(\vc r)$ for (a) LS and (b) TJ packings at system sizes of $N=10000$ and $N=2000$, respectively.  Data for positive $c$ is shown using dashed lines of the same color in (b).  The thick black dashed guide line shows a slope of $-1/r^2$.}
    \label{fig:crDisorderedLog}
  \end{centering}
\end{figure*}

The difference in the protocols' approaches to jamming may be readily traced back to differences in the dynamics involved: on one hand, the constant thermal motion inherent to LS acts to equilibrate packing and avoid metastable branches terminating at low-density jammed states; an aggressive expansion rate works against this, though one must worry about the algorithm becoming trapped in an unstable mechanical equilibrium (which is, by definition, not a jammed state). 
The possible displacements obtained by TJ are highly degenerate since, in general, there are many different displacements which allow for the same increase in the packing fraction (which is limited to a small value so that the linear approximations in the LPs' formulation remain reasonably accurate). 
Therefore, TJ tends to displace spheres in a ``lazy'' fashion, only moving what is necessary to increase the packing fraction and no more.

Evidence of this aforementioned qualitative difference may be observed directly; Figure \ref{fig:visual} shows snapshots of two-dimensional packings of monodisperse disks created by TJ and LS (using a rapid compression rate) at a packing fraction of $\phi=0.55$. 
In two dimensions just as in three, we can see that the structures of the packings are qualitatively different. 
In particular, the TJ algorithm exhibits clustering of particles that might be quickly dispersed through thermal motion; in the absence of this, the clusters continue to combine and aggregate as jamming is approached.
We point out that particles within these clusters do not necessarily contact one another; some separation is expected to remain between particles owing to the nonlinearities that are not captured in the TJ algorithm's linear approximation to the packing problem.
In LS, on the other hand, particles tend to space themselves out more uniformly through their thermal motion.
We note in passing that, while this difference does not prevent the LS algorithm from discovering MRJ-like states in 3D, the two-dimensional case was recently shown to be considerably more subtle---the difference in how TJ creates jammed packings has led to the first observations of MRJ-like packings of monodisperse disks in two dimensions, whereas the LS algorithm and other standard protocols are unable to observe them, finding significantly more ordered, polycrystalline structures even under rapid compression \cite{Atkinson_2014}.

%
% TJ:
%  color: 0,0.3,0.7
%  Source stored in ../Figures/Visual/TJ100.dat
% LS:
%  color: 0.7,0,0
%  Source stored in ../Figures/Visual/LS100.dat
%
% N=1000 versions available, too
%
\begin{figure}[hbt]
  \begin{centering}
    $\begin{array}{ccc}
      \includegraphics[width=0.22\textwidth,clip]{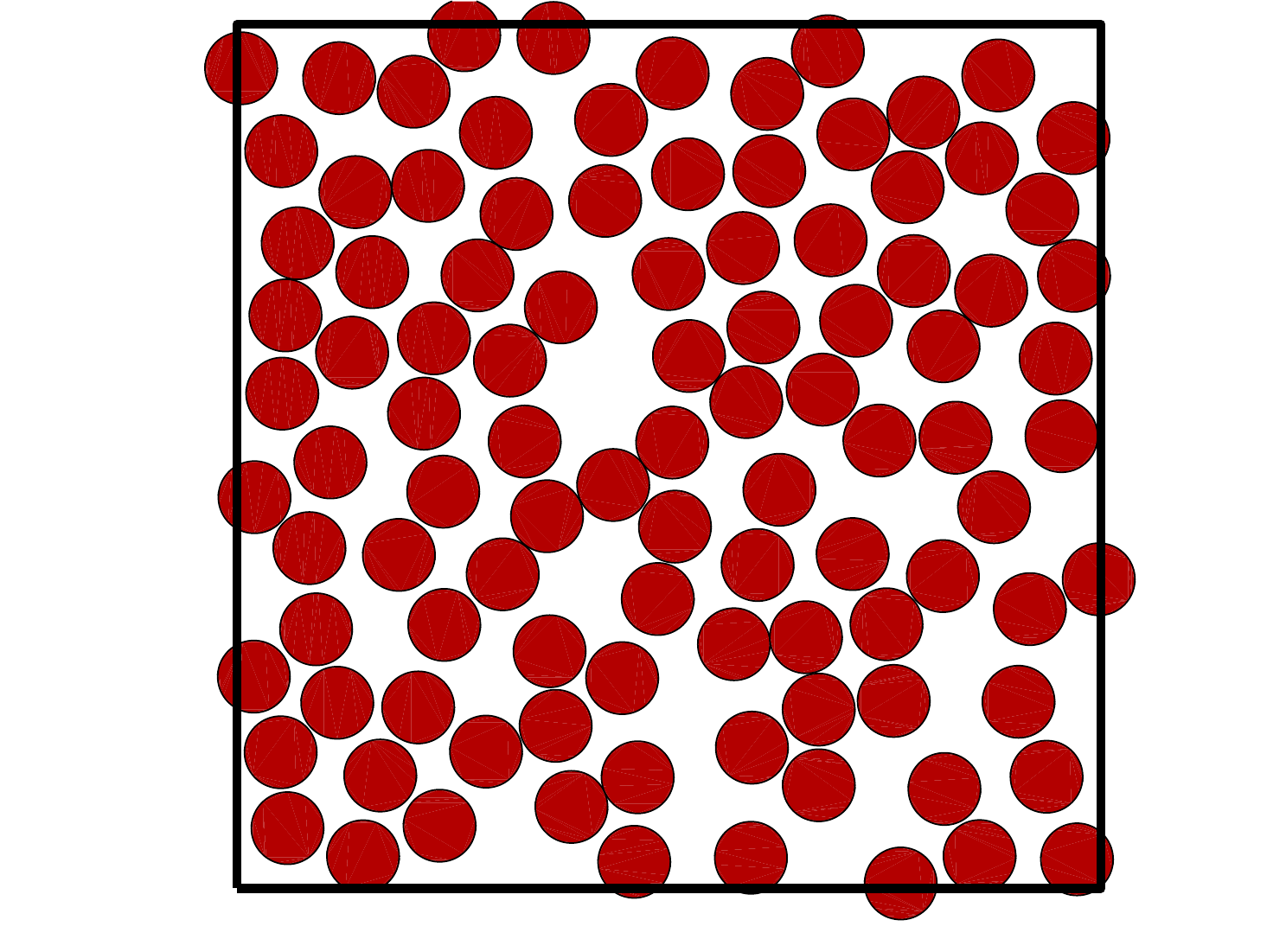}
      &
      \includegraphics[width=0.22\textwidth,clip]{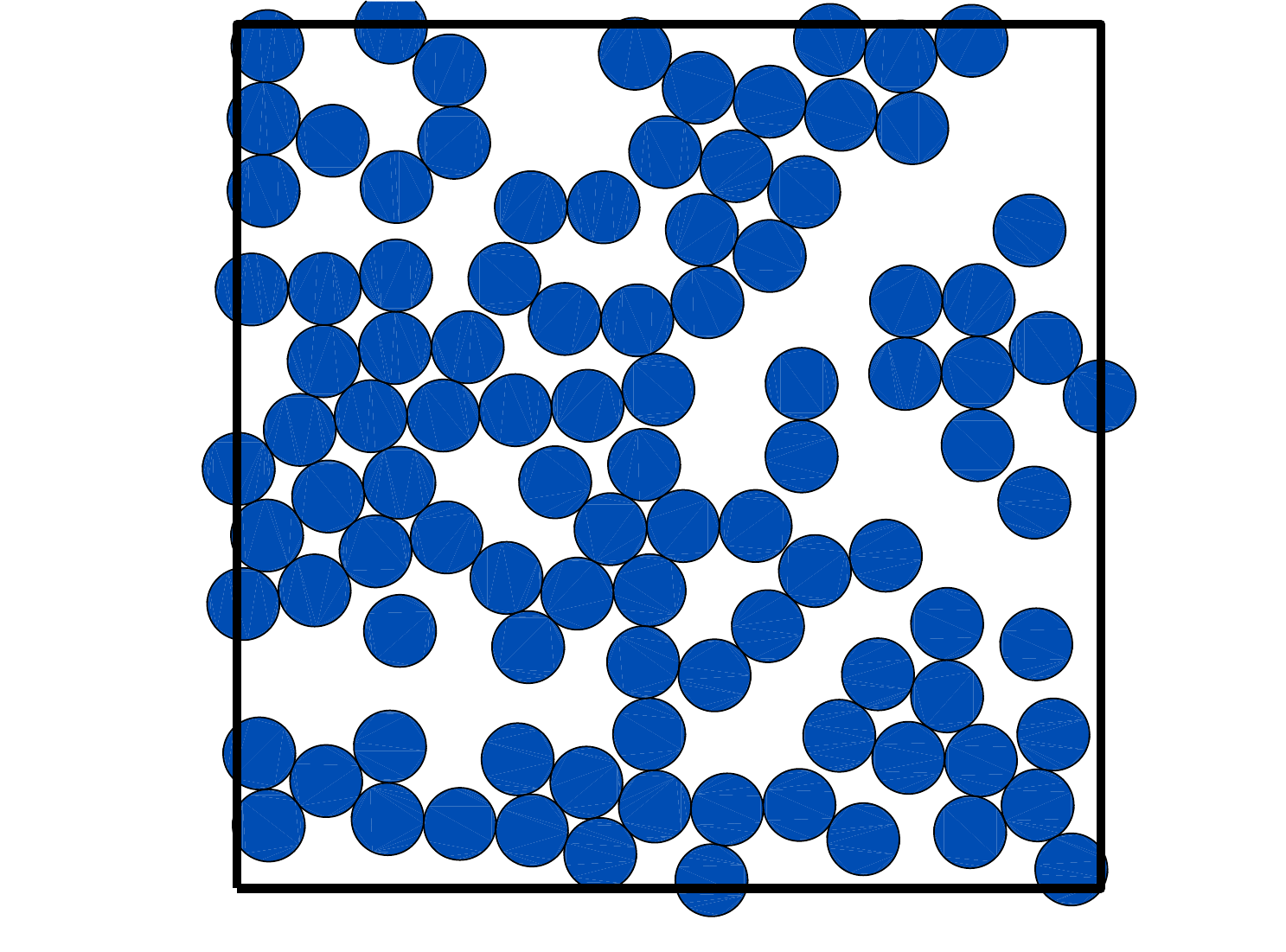}
      \\
      \mbox{(a)}
      &
      \mbox{(b)}
    \end{array}$
    \caption{(Color online.) Two-dimensional packings of monodisperse disks created by the (a) LS and (b) TJ algorithms at a packing fraction of $\phi=0.55$.}
    \label{fig:visual}
  \end{centering}
\end{figure}

%%%%%%%%%%%%%%%%%%%%%%%%%%%%%%%
%%        Discussion         %%
%%%%%%%%%%%%%%%%%%%%%%%%%%%%%%%

\section{Conclusions and Discussion}
\label{sec:Conclusion}

%Recap
In this work, we have compared and contrasted the approach of both ordered and disordered hard-sphere packings towards jammed states through considering the behavior of their structure factors and direct correlation functions.
By considering the degree and position of singularities in $c(\vc r)$ as well as how they are changed by the convolutions found in Eq.\ (\ref{eqn:OZ}), we have established quantitative statements about the structure of the direct correlation function with regards to features it inherits from $g_2(\vc r)$.
These relations provide a concrete means of identifying what features must be expressed in $c(\vc r)$ if one hopes to reproduce various details in $g_2(\vc r)$ accurately.

%TJ vs LS and tau
Moreover, we found that the LS and TJ protocols approach their respective jammed states in markedly different manners, as shown by various pair statistics.
Specifically, the structure factor of TJ-generated packings shows anomalous increasing behavior for small $k$ at intermediate densities, and generally remains closer to $S=1$ at all densities leading up to jamming when compared to LS.
The order metric $\tau$ compares a configuration's pair statistics to that of an uncorrelated (Poisson) point process, which may be thought of as a maximally disordered reference state.
In this sense, $\tau$ may be thought of as a ``disorder metric''.
At low to intermediate densities, $\tau$ suggests that packings created by TJ are more disordered on large length scales, but more ordered on short length scales as evidenced by the crossover as the truncation in the integration domain $k_{max}$ is made increasingly large.
This is consistent with the intuition that TJ does not disturb the packings as they are compressed as much as LS does, leaving the large-scale characteristics similar to the initial conditions obtained from low-density RSA.
On the other hand, the formation of near-contacts well before jamming may be interpreted as a sort of ``ordering'' that LS avoids through equilibration;
therefore, LS yields configurations that are more disordered locally at densities far from jamming.

TJ shows signs of particles in close proximity at surprisingly low densities as evidenced by the appearance of a clear precursor to the delta function at $c(1)$ and corresponding $-1/r$ scaling within the core as $r \rightarrow 0$.
We have shown that the latter numerical observation can be predicted from theoretical considerations using a Yukawa model for $c(\vc r)$.
By evaluating $\tau$, we see that these near-contacts that cause the delta function to appear are distributed across a range of pair separations, and the delta function's precursor is not ``sharp'' until higher densities.
This is to be expected because of the linear approximations that TJ makes as the packing is compressed; the inaccuracies due to nonlinear contributions are largest when large changes in the system configuration (particle translations and box deformations) are made.
This is the case at densities far from jamming, where the linear approximations to the packing problem still leave a large amount of configuration space accessible.
Nonetheless, this feature in the intermediate-density structures produced by the TJ algorithm suggests that it possesses important qualitative commonalities with the physical process of compressing hard-spheres embedded in a dampening background, providing a conceptual physical analog to the algorithm as witnessed in practice.

%Confirmation of 1/r^2
It has been suggested previously \cite{Torquato_2003_Hyperuniform,Hopkins_2012} that the hyperuniform, linear nonanalytic behavior of $S(\vc k)$ for MRJ-like packings ought to give rise to a long-ranged direct correlation function which exhibits a power-law decay of $c \propto -1/r^2$. 
We have confirmed this numerically using simulated packings of hard-spheres generated by two very different protocols, adding to the evidence to the conjectured link between jamming and hyperuniformity \cite{Donev_2005_Unexpected,Torquato_2010} and supporting the idea that the emergence of large-$r$ scaling behavior consistent with hyperuniformity may be regarded as a structural precursor to jamming. 
It would be interesting to consider new semi-empirical forms for $c(\vc r)$ incorporating this long-range behavior in order to gain understanding regarding its structural consequences.

%FCC, k^2, and percolation.
It is interesting that the structure factor of the FCC crystal exhibits scaling that is constant for low $k$, but gives way to $k^2$ beginning at wavenumbers on the order of unity, extending to an increasingly large maximum wavenumber as jamming is approached. 
We noted above that this would imply that the average pair distance between any given pair of nearest-neighbor particles might be spatially uncorrelated to a good approximation. 
It has been observed elsewhere \cite{Ikeda_2015} that the fluctuating component of the structure factor in disordered packings of thermally-excited soft-spheres exhibits a similar quadratic scaling at densities slightly above the jamming transition density. 
We have noticed that this behavior is also exhibited for disordered hard-sphere packings at packing fractions below, but close to jamming, suggesting similarly that the pair separations between nearest-neighbor particles may fluctuate in an uncorrelated manner to a good approximation.

%Next paper
We suspect that the aforementioned differences between the LS and TJ algorithms should be evident in other ways. 
In a subsequent paper \cite{Atkinson_2016_Perc}, we will study our hard-sphere systems in the context of two different percolation problems. 
In the first, we decorate the hard cores with a perfectly penetrable shell (this is known as the ``cherrypit model'' \cite{Torquato_2002}) and tuning the size of this shell for various configurations to explore percolation criticality. 
Based on our findings here, one would expect for TJ that the critical shell thickness would become very small rather quickly, whereas it might decrease more steadily towards zero for LS as jamming is approached.  Moreover, one might expect to find structural differences in the percolating clusters between these two algorithms. 
Second, we investigate percolation by measuring the time-averaged magnitudes of pair forces in nearly-jammed, structurally-arrested configurations as a function of a minimum ``threshold'' force. 
This approach serves to ``average out'' fluctuations in particle position, and thus provides insight into the role of fluctuations in the incipient structure of disordered, jammed systems. 
Both approaches provide insight into the jamming process using static structural features, expanding upon the work presented here.

\begin{acknowledgments}
The authors thank Yang Jiao and Adam Hopkins for their careful reading of this manuscript and valuable comments.
This work was supported in part by the National Science Foundation under Grant No.\ DMS-1211087.
\end{acknowledgments}

\ifshowappendix

\appendix

\section{Numerical procedure for obtaining $c(r)$ from $S(k)$ data}
\label{sec:appendix_crNumerics}

In this appendix, we provide procedural details regarding our numerical computation of the direct correlation function used to ascertain the large-$r$ behavior.

We begin by measuring $S(\vc k)$ at wavevectors that are integer combinations of the columns of the reciprocal matrix $\mtx \Lambda_R$ defined as $\mtx \Lambda_R = [(2\pi) \mtx \Lambda^{-1}]^T$, where the columns of $\mtx \Lambda \in \reals^{d \times d}$ span the fundamental cell of our system in direct space.  Measurements are then binned and averaged according to their wavenumber with a bin width of $\Delta k/2\pi = 0.01$, so that we have reported measurements at wavenumbers given by $k_n = (\Delta k/2\pi)(1/2+n)$ for $n=0,1,2\dots$.  Because the density of wavevectors scales as $k^{d-1}$, we randomly select wavevectors with a probability proportional to the inverse of this density such that, for higher wavenumbers, the expected number of measurements per bin is $E(n_s) = 10^5$.  At smaller wavenumbers, the structure factor is measured at every available wavevector.  

Once we have obtained data for all of our packings, we perform an ensemble average.  If there are empty bins, then we linearly interpolate a value for those bins, expecting that this is representative of the large-system limit.  We also linearly extrapolate $S(k)$ down to zero if any bins are missing data; our results are qualitatively robust against small variations in this extrapolation.

%At high densities, we observe that our data is well-fit by an asymptotic form of $S(\vc k) = 1 + A \sin(k)/k$, reflecting the (expected) presence of a delta function at $c(r=1)$.  We fit a value for $A$ to our data for $k/2\pi > 14$ and append this \red{needed? NOPE! :)}

Our data is then converted via Eq.\ (\ref{eqn:ck}) to give us $\tilde c(\vc k)$.  Because of the asymptotic behavior of $\tilde c(\vc k)$, we find that it is necessary to apply a convolution in order to eliminate artifacts caused by difficulties associated with numerically integrating this high-frequency behavior.  We do this by multiplying $\tilde c(\vc k)$ with the Fourier transform of a triangular window in direct space given by $w(\vc r) = (3/\pi r_c^3)(1-r/r_c) (1-\Theta(r-r_c))$.  The Fourier transform of this radial function is 
\begin{equation}
  \tilde w(\vc k) = 
  \begin{cases}
    \frac{12}{(k r_c)^4} \left( 2-2 \cos(k r_c) - k r_c \sin(k r_c) \right), & k>0
    \\
    1, & k=0 
  \end{cases}.
\end{equation}

For a general three-dimensional, radial function $f(\vc r) \equiv f(\norm{\vc r})$, the Fourier transform and its inverse may be expressed \cite{Torquato_2002} as
\begin{eqnarray}
  \tilde f(\vc k) &=& \begin{cases}
    \frac{4\pi}{k} \int_0^\infty r f(\vc r) \sin(k r) dr, & k>0
    \\
    \int_{\reals^3} f(\vc r) d \vc r, & k=0
  \end{cases}
  \label{eqn:FT3D}
  \\
  f(\vc r) &=& \begin{cases}
    \frac{1}{2 \pi^2 r} \int_0^\infty k \tilde f(\vc k) \sin(k r) dk, & r>0
    \\
    \frac{1}{(2 \pi)^3}\int_{\reals^3} \tilde f(\vc k) d \vc k, & r=0
  \end{cases}.
  \label{eqn:IFT3D}
\end{eqnarray}
Note that both the forward and inverse transforms are equivalent up to their scaling coefficients.  In order to take advantage of the usual one-dimensional fast Fourier transform algorithm for our three-dimensional, radial $\tilde c(\vc k)$, we take our discrete $C_n = \tilde c(n \Delta k)$ for $n=0,1,2,\dots,l$ and compute 
\begin{equation}
  C'_n = \begin{cases}
    (n-l) \Delta k C_{l-n-1}, & n=0,1,\dots,l
    \\
    0, & n=l+1
    \\
    (n-l-1) \Delta k C_{n-l-1}, & n=l+2,\dots,2l+1
  \end{cases}.
\end{equation}
We then compute the usual one-dimensional inverse FFT on this data, defined here as
\begin{eqnarray}
c'_m &=& \mc F^{-1} [C'_n;m]
\nonumber
\\
 &=& \frac{\Delta k}{2l+1} \sum_{n=0}^{2l+1} C'_n e^{2 \pi n m/(2l+1)}
\label{eqn:IFFT}
\end{eqnarray}
where $m=1,\dots,2l+1$.  We then apply the prefactor and a phase correction because to correct for the fact that the index $n=l+1$ corresponds to $k=0$ to obtain
\begin{equation}
c_m = \frac{c'_m}{2 \pi^2 r_m} e^{-2 \pi i m l/(2l+1)}
\end{equation}
where $r_m = m \Delta r$ and $\Delta r = 2\pi/\Delta k (2l+1)$.  Through analogy with Eq.\ (\ref{eqn:IFT3D}), one expects that $c_m$ is completely imaginary, while one expects the Fourier transform to be completely real-valued.  This is reconciled by dropping the imaginary unit from $c_m$; doing so is justified since the imaginary prefactor is expected if one applies Euler's formula to the exponential term in Eq.\ (\ref{eqn:IFFT}), but is missing as a prefactor to the sine term in Eq.\ (\ref{eqn:IFT3D}).  Finally, the value for $c_0$ corresponding to $c(r=0)$ is obtained by integrating $C_n$ numerically according to Eq.\ (\ref{eqn:IFT3D}).

Once the direct space $c(r)$ has been found (represented discretely through $c_m$), one must then be sure to truncate the data at $r_{max}=L_{max}/2$ where $L_{max}$ is the width of the simulation box; data beyond this point is subject to finite-size artifacts.

%\begin{itemize}
%\item Average the $S(k)$ data.
%\item InterpolateTo() a k-series that starts on 0 instead of 1/2
%\item Extrapolate the low-k data down to zero
%\item cut off the data around $kD/2p=14$ or so (wherever it's still not noisy), and fit on a "tail" of the form "$1 + A/k'*sin(2*pi*k')$" [where $k' = kD/2p$]; extend that out to $kD/2p=300$ or so (why not?)
%\item Make a Fourier-transformed triangular convolution window $w(k)$
%\item Compute $1+(S(k)-1)*w(k)$
%\item Compute $c(k)$, then $c(r)$
%\item find the r-cutoff for the system size $L/2 = 1/2*((N*pi)/(6*phi))^(1/3)$ and cut the $c(r)$ there.
%\end{itemize}

\fi %showappendix

\ifusebibtex

% EXTRA citation just to get it into the bbl:
\cite{Percus_1964}.

\bibliography{paper}

\else %!usebibtex

\fi %usebibtex
\fi %showpaper

\end{document}